\documentclass[prl,twocolumn,superscriptaddress,longbibliography,preprintnumbers,floatfix,nofootinbib]{revtex4-1}

\usepackage{url}
\usepackage{xspace}
\usepackage{dsfont}
\usepackage{amssymb}
\usepackage{amsmath}
\usepackage{graphicx}
\usepackage[caption=false]{subfig}
\usepackage[colorlinks=true,citecolor=blue]{hyperref}
\usepackage{multirow}
\usepackage{hhline}

\usepackage{color}
\definecolor{mit-red}{rgb}{0.64,.12,0.2}
\definecolor{darkred}{rgb}{1.0,0.1,0.1}
\definecolor{darkgreen}{rgb}{0.1,0.7,0.1}
\definecolor{darkblue}{rgb}{0.1,0.1,1.0}

\usepackage{amsmath}
\DeclareMathOperator*{\argmax}{argmax}

\DeclareRobustCommand{\Tab}[1]{Table~\ref{tab:#1}}

\DeclareRobustCommand{\Fig}[1]{Fig.~\ref{fig:#1}}

\DeclareRobustCommand{\Eq}[1]{Eq.~(\ref{eq:#1})}
\DeclareRobustCommand{\Eqs}[2]{Eqs.~(\ref{eq:#1}) and (\ref{eq:#2})}
\DeclareRobustCommand{\Reference}[1]{Ref.~\cite{#1}}

\newcommand{\GaussianAnsatz}{{Gaussian Ansatz}\xspace}

\newcommand{\Adam}{{\sc Adam}\xspace}

\begin{document}

\title{Learning Uncertainties the Frequentist Way:\\  Calibration and Correlation in High Energy Physics }

\preprint{MIT-CTP 5431}

\author{Rikab Gambhir}
\email{rikab@mit.edu}
\affiliation{Center for Theoretical Physics, Massachusetts Institute of Technology, Cambridge, MA 02139, USA}
\affiliation{The NSF AI Institute for Artificial Intelligence and Fundamental Interactions}

\author{Benjamin Nachman}
\email{bpnachman@lbl.gov}
\affiliation{Physics Division, Lawrence Berkeley National Laboratory, Berkeley, CA 94720, USA}
\affiliation{Berkeley Institute for Data Science, University of California, Berkeley, CA 94720, USA}

\author{Jesse Thaler}
\email{jthaler@mit.edu}
\affiliation{Center for Theoretical Physics, Massachusetts Institute of Technology, Cambridge, MA 02139, USA}

\affiliation{The NSF AI Institute for Artificial Intelligence and Fundamental Interactions}
\begin{abstract}
Calibration is a common experimental physics problem, whose goal is to infer the value and uncertainty of an unobservable quantity $Z$ given a measured quantity $X$.
Additionally, one would like to quantify the extent to which $X$ and $Z$ are correlated.
In this paper, we present a machine learning framework for performing frequentist maximum likelihood inference with Gaussian uncertainty estimation, which also quantifies the mutual information between the unobservable and measured quantities.
This framework uses the Donsker-Varadhan representation of the Kullback-Leibler divergence---parametrized with a novel \GaussianAnsatz---to enable a simultaneous extraction of the maximum likelihood values, uncertainties, and mutual information in a single training.
We demonstrate our framework by extracting jet energy corrections and resolution factors from a simulation of the CMS detector at the Large Hadron Collider.
By leveraging the high-dimensional feature space inside jets, we improve upon the nominal CMS jet resolution by upward of 15\%.
\end{abstract}

\maketitle


One of the most foundational tasks in high energy physics (HEP) is the inference of an unobservable quantity given a  measured quantity, which is often referred to as \textit{calibration}.
For example, the kinematic properties of a given particle must be reconstructed from signatures registered in various detector elements.
This inference task can be challenging when the reconstruction requires high-dimensional inputs.
Machine learning (ML) is a natural tool for performing high-dimensional reconstruction, and there has been significant progress in utilizing ML methods for estimating the energies of various objects, including
photons~\cite{CMS:2020uim},
muons~\cite{Kieseler:2021jxc},
single hadrons~\cite{Belayneh:2019vyx,ATL-PHYS-PUB-2020-018,Akchurin:2021afn,Akchurin:2021ahx,Polson:2021kvr,Pata:2021oez},
and sprays of hadrons (jets)~\cite{ATL-PHYS-PUB-2018-013,ATL-PHYS-PUB-2020-001,CMS:2019uxx,Haake:2018hqn,Haake:2019pqd,Baldi:2020hjm,Komiske:2017ubm,ATL-PHYS-PUB-2019-028,Maier:2021ymx,Kasieczka:2020vlh,ArjonaMartinez:2018eah, jec_with_gnn_regression} at colliders;
kinematic reconstruction in deep inelastic scattering~\cite{Diefenthaler:2021rdj,Arratia:2021tsq};
and neutrino energies in a variety of experiments~\cite{Liu:2020pzv,EXO:2018bpx,Baldi:2018qhe,Abbasi:2021ryj,IceCube:2020yct,Carloni:2021zbc}.
Further ideas can be found in \Reference{Feickert:2021ajf}.

Abstractly, the calibration task can be described as quantifying the relationship between two random variables $X\in\mathbb{R}^M$ and $Z\in\mathbb{R}^N$.
Here, $X$ is the measured quantity and $Z$ is the unobservable (``latent'') quantity.%
\footnote{Throughout this paper, upper case letters represent random variables and lower case letters represent realizations of those random variables.}
A reconstruction technique produces a function $\hat{z}:\mathbb{R}^M\rightarrow\mathbb{R}^N$, which is determined by minimizing a loss functional over sample data (real or synthetic).
While ML methods are effective even when $M$ and $N$ are large, most existing methods have the undesirable property of being prior dependent~\cite{priordependence}.
This means that $\hat{z}$ depends on the probability density $p(z)$ used during training.
As a result, the calibration is not universal and caution must be taken when applying it to different event samples.

Furthermore, some calibration methods simply produce a \textit{point estimate}, with no estimation of the corresponding uncertainty.
In the HEP context, this uncertainty is usually called the \textit{resolution}.
Quantifying the reconstruction resolution is relevant for a variety of purposes, including
the computation of significance variables~\cite{CMS:2019ctu,Nachman:2013bia} and
background estimation~\cite{ATLAS:2012qgw,ATLAS:2021kxv}.
Various ML approaches for resolution determination have been recently studied for HEP~\cite{Sirunyan:2019wwa,Cheong:2019upg,Bollweg:2019skg,Bellagente:2021yyh,1806026,Araz:2021wqm,Kronheim:2021hdb}, but they typically require additional training or model complexity.
See \Reference{https://doi.org/10.48550/arxiv.2107.03920} for a complementary approach to frequentist inference.

In this paper, we introduce a simple ML framework for calibration that simultaneously estimates the following quantities:
\begin{enumerate}
    \item A prior-independent maximum-likelihood calibration, $\hat{z}(x)=\argmax_z p(x|z)$;
    \item A Gaussian resolution around $\hat{z}(x)$, $\hat{\sigma}_z(x)$;
    \item The log-likelihood ratio, $\log \frac{p(x|z)}{p(x)}$; and
    \item The mutual information between $X$ and $Z$, $I(X;Z)$.
\end{enumerate}
To extract $\hat{z}(x)$ and $\hat{\sigma}_z(x)$ in a single training, we use a novel \textit{\GaussianAnsatz} to parametrize the log-likelihood ratio with an interpretable network architecture.
Mutual information is a powerful statistic for quantifying the (non-linear) correlation between two random variables, and it appears due to our choice of loss function.
After describing the \GaussianAnsatz construction, we illustrate the above features in a case study involving jet reconstruction at the Large Hadron Collider (LHC).

Our calibration method builds upon the  Mutual Information Neural Estimator (MINE) introduced in \Reference{belghazi2018mine}.
With MINE, the Donsker-Varadhan representation~\cite{Donsker1975AsymptoticEO} of the Kullback-Leibler divergence~\cite{kullback1951information} is used to estimate $I(X;Z)$ by training a network to minimize a particular loss functional.
We first show that a network minimizing this loss functional yields the likelihood $p(x|z)$, which in principle contains all the information necessary for frequentist inference.
Performing this inference in practice, though, involves difficult optimization tasks, which are even more difficult if one wants to extract the resolution.
With the \GaussianAnsatz, we parametrize the MINE network such that the inferred value, and especially the resolution, are easy to extract after ML training.

The starting point for our calibration method is the concept of mutual information (MI), defined as:
\begin{align}
    \label{eq:MI_definition}
    I(X;Z) = \int\text{d}x\,\text{d}z\, p(x,z) \log\frac{p(x,z)}{p(x)\,p(z)} ,
\end{align}
where $p$ denotes the probability density of the respective random variable.
This equation has the property that $I(X;Z)=0$ if and only if $X$ and $Z$ are independent, which is equivalent to $p(x,z)=p(x)\, p(z)$.
Therefore, the MI quantifies the interdependence between $X$ and $Z$.

The MI is a special case of the Kullback-Leibler (KL) divergence, $D_{\text{KL}}(P||Q)$, when $P=P_{XZ}$ is the joint probability distribution of $X$ and $Z$ (i.e.~$p(x,z)$), and $Q=P_X\otimes P_Z$ is the product of the marginals (i.e.~$p(x)\,p(z)$).
It is well known that the KL divergence can be cast in the Donsker-Varadhan (DV) representation~\cite{e5879cd3d84b462abf51f06791e5ba28}:
\begin{align}
    D_{\text{KL}}(P||Q) = \sup_{T \in \mathcal{T}}\Big\{\mathbb{E}_P\left[T\right] - \log\left( \mathbb{E}_Q\left[e^T\right]  \right)   \Big\}\,,
\end{align}
where $\mathbb{E}_\bullet$ represents the expectation value over probability density $\bullet$, and the supremum is over the space of functions $\mathcal{T}$ such that both expectations are finite.

Following the MINE construction in \Reference{belghazi2018mine}, we use the DV representation to build an estimator for the mutual information from a finite dataset.
For functions $T:\mathbb{R}^M \times \mathbb{R}^N \to \mathbb{R}$, we can place a lower bound on $I(X;Z)$ by minimizing a loss functional $\mathcal{L}_{\rm DVR}$ over $T \in \mathcal{T}$:
\begin{align}
\label{eq:mine}
    I(X;Z) \geq - \inf_{T\in\mathcal{T}} \mathcal{L}_{\rm DVR}[T]\,,
\end{align}
where the DV representation (DVR) loss is:
\begin{equation}
    \label{eq:DV_loss}
    \mathcal{L}_{\rm DVR}[T] = -\Big(\mathbb{E}_{P_{XZ}}\left[T\right] - \log\left( \mathbb{E}_{P_X\otimes P_Z}\left[e^T\right]  \right) \Big)\,.
\end{equation}
Given a finite dataset of $(x,z)$ pairs, the expectations in \Eq{DV_loss} can be estimated from sample averages.
To estimate the second term, one can simply shuffle the $x$'s and $z$'s, as done in \Reference{belghazi2018mine}.
The space of functions $\mathcal{T}$ can be parametrized by neural networks, in which case the DVR loss functional can be minimized using standard gradient descent.
As long as $\mathcal{T}$ is sufficiently expressive, the bound in \Eq{mine} will be saturated, so the minimum loss is an estimate of $-I(X;Z)$.%
\footnote{
Other loss functionals exist that are capable of providing lower bounds on MI.
For example, if we write the $f$-divergence representation of the KL divergence~\cite{nowozin2016fgan,Nguyen_2010}, the corresponding loss functional is a variation of the maximum likelihood classifier (MLC) loss \cite{DAgnolo:2018cun,DAgnolo:2019vbw,nachman2021e}:
\begin{align}
    \mathcal{L}_{\rm MLC}[T] = - \Big(\mathbb{E}_{P_{XZ}}\left[T \right] -  \mathbb{E}_{P_X\otimes P_Z}\left[e^T - 1\right]\Big).
\end{align}
Numerical  and analytic studies \Reference{belghazi2018mine, ruderman2012tighter}, as well as our own empirical studies, show that the DVR loss has better numerical convergence properties than the MLC loss.}

Taking the functional derivative of the DVR loss functional with respect to $T$, we see that the supremum of $\mathcal{L}[T]$ is obtained when:
\begin{align}
    \label{eq:T_result}
    T(x,z) =  \log\frac{p(x|z)}{p(x)} + c\,,
\end{align}
where $c$ is any constant that we can set to zero without loss of generality.%
\footnote{In practice, we determine and then subtract $c$ numerically by noting that the second term of \Eq{DV_loss} is an estimate of $c$ in the asymptotic limit.}
Therefore, if the MINE is well trained, we can use $T$ as an estimate of the log-likelihood density ratio.
As with most machine learning applications, this requires that the space of neural networks $\mathcal{T}$ is sufficiently expressive, that there is enough training data, and that the gradient descent algorithm successfully finds the minimum of \Eq{DV_loss}.
Given this, we can then perform maximum likelihood inference given $x$:
\begin{align}
    \hat{z}(x) &= \argmax_{z}\,T(x,z). \label{eq:argmax}
\end{align}
Crucially, this inference strategy for $z$ is independent of the prior $p(z)$, which is a property desirable for calibration tasks.
Unlike for standard regression~\cite{priordependence}, the learned estimate $\hat{z}$ does not depend on the distribution of $z$ samples in the training set.%
\footnote{If desired, one could do Bayesian inference and obtain the posterior $p(z|x)$ by adding the prior $\log p(z)$ to $T(x,z)$.}

If $X$ does not contain complete information about $Z$, then there will be uncertainty in our inference of $z$.
Assuming the likelihood $p(x|z)$ is well approximated by a Gaussian density, the uncertainty in the inference is given by the covariance matrix:
\begin{align}
    \left[\hat{\sigma}_{z}^2(x)\right]_{ij} = - \left[ \frac{\partial^2 T(x,z)}{\partial z_i \, \partial z_j}  \right]^{-1}\biggr\rvert_{z=\hat{z} }\,, \label{eq:second_derivative}
\end{align}
which is again prior independent.

So far, we have shown that the MINE network can be used to perform frequentist inference. While $T$ itself depends on the prior $p(z)$, the inference $\hat{z}$ and resolution $\hat{\sigma}_z$ do not.
However, both the maximum likelihood estimate in \Eq{argmax} and the local resolution in \Eq{second_derivative} are difficult to evaluate numerically.
In the case of maximization, the learned $T$ may be highly non-convex and the true maxima difficult to find using gradient descent.
In the case of the second derivative, its evaluation is numerically sensitive to the choice of activation function in the MINE network.
For example, if one uses the common Rectified Linear Unit (ReLU) activation function or its variants, then all analytic second derivatives of the network are zero.

In order to facilitate a numerical estimate of the maximum likelihood and local resolution, we introduce the following \GaussianAnsatz parametrization for $T$:
\begin{align}
\label{eq:gaussian_ansatz}
    T(x,z) &= A(x) + \big(z-B(x)\big)\cdot D(x) \nonumber\\&\quad + \frac{1}{2} \big(z-B(x)\big)^T \cdot C(x,z) \cdot \big(z-B(x)\big)\,,
\end{align}
where $A:\mathbb{R}^N\xrightarrow{}\mathbb{R}$ , $B:\mathbb{R}^N\xrightarrow{}\mathbb{R}^M$,  $C:\mathbb{R}^N\times\mathbb{R}^M\xrightarrow{}\text{Sym}(M, \mathbb{R})$, and  $D:\mathbb{R}^N\xrightarrow{}\mathbb{R}^M$ are each neural networks.
We call this the \GaussianAnsatz, since it resembles the logarithm of a Gaussian likelihood density.
Unlike a Gaussian likelihood, though, the \GaussianAnsatz is highly expressive, and is in fact a universal function approximator.
Specifically, any function $f(x,z)$ that admits a Taylor expansion in $z$ around $B(x)$ can be expanded in this form.
The functions $A(x)$ and $D(x)$ capture the zeroth and first order dependencies of $f$ on $z$, respectively.
The function $C(x,z)$ captures any quadratic or higher dependence of the Taylor expansion of $f$ on $z$.

The \GaussianAnsatz enables an elegant strategy to extract \Eqs{argmax}{second_derivative}.
Since the optimal $T(x,z)$ is bounded from above, we can take $D(x)$ to be everywhere zero without loss of expressivity.
In this case, $T$ will achieve critical values at $z = B(x)$.
Moreover, if $C(x, B(x)) < 0$, then these critical values will be (local) likelihood maxima:
\begin{equation}
        \hat{z}(x)  = B(x).     \label{eq:y_ML} \\
\end{equation}
While the \GaussianAnsatz does not necessarily protect against local maxima, it does yield a numerical estimate of the local resolution:
\begin{equation}
    \hat{\sigma}_{z}^2(x)   = - \big[C(x,B(x))\big]^{-1}. \label{eq:cov_ML}
\end{equation}
Moreover, the (negative) loss of the \GaussianAnsatz with respect to the functional in \Eq{DV_loss} will be a lower bound for the mutual information $I(X;Z)$, which is saturated in the asymptotic limit of an infinitely large network with infinite data.

The \GaussianAnsatz is therefore capable of estimating---from a single dataset of $(x,z)$ pairs and a single training---the maximum likelihood inferred value of $z$ given $x$, the local resolution on that inference, and the mutual information between $X$ and $Z$.
This can be achieved without having to perform any additional optimization problems, derivative estimations, or postprocessing beyond the single matrix inversion in \Eq{cov_ML}.
In practice, we find it convenient to start the training with non-zero $D(x)$ to aid the convergence of the model, and then numerically force $D \to 0$ through an increasing $L_1$ regularization.
This helps the model achieve a global, rather than local, minimum.

We now demonstrate the \GaussianAnsatz on an experimental collider physics task: determining jet energy corrections (JECs) and resolutions (JERs)~\cite{Khachatryan_2017}.
(At the LHC, one typically calibrates transverse momenta $p_T$ instead of energies, but the terms JECs and JERs are still used.)
Jets are collimated sprays of particles that are produced ubiquitously in high-energy collisions.
One does not have access to the ``true'' jet energy, however, because its constituent particles are filtered through a complicated and nonlinear detector response.

Assuming one has a good detector model, though, one can \emph{generate} truth-level quantities (GEN, corresponding to $Z$) and then \emph{simulate} the detector response (SIM, corresponding to $X$).
Performing a \emph{simulation-based calibration}, one can infer the ``true'' jet energy from a set of measured particle momenta in a jet.
The multiplicative JEC factor is then defined such that the inferred jet momenta is:
\begin{equation}
\label{eq:JEC_def}
\hat{p}_{T} \equiv \text{JEC} \times p_{T,\text{SIM}}\approx p_{T,\text{GEN}}.
\end{equation}
JEC factors are often further refined through a \emph{data-based calibration} using well-understood control samples, though this is separate from the procedure considered here.
The JER factor arises because the inferred and generated $p_T$ values in \Eq{JEC_def} are not identical.
The JER is typically expressed as a fractional quantity:
\begin{equation}
\label{eq:JER_def}
\hat{\sigma}_{p_{T}} = \text{JER} \times p_{T,\text{SIM}}.
\end{equation}
In the language of statistics, the JER is a type of ``uncertainty'', since it represents the limited information about $Z$ contained in $X$.
In the HEP context, though, this quantity is instead called a ``resolution''; see \Reference{priordependence} for further discussion.

The JEC factor is a function of the measured quantities, primarily the detector-level jet $p_T$ and pseudorapidity $\eta$.
The JEC can be obtained from fits to simulation \cite{2011jes,CMS:2016lmd, ATLAS:2017bje,ATLAS:2014hvo,ATLAS:2019oxp} using a technique called numerical inversion~\cite{Cukierman_2017}.
The JER can be also determined in simulation by fitting the peak region of the detector response $\hat{p}_T/p_{T,\text{SIM}}$ to a Gaussian distribution.  
Here, we consider an alternate (and arguably simpler) approach to JEC and JER extraction.

For our case study, we use the \GaussianAnsatz to improve upon the JEC factors provided by the CMS experiment in their 2011 public data release~\cite{cmspressrelease}.
We use the same 2011 CMS Open Simulation~\cite{cernopendata} samples as in \Reference{Komiske_2020}, which are based on dijets generated in \textsc{Pythia 6}~\cite{Sj_strand_2006} with a \textsc{Geant4}-based~\cite{AGOSTINELLI2003250} simulation of the CMS detector.
This dataset was translated from the original CMS AOD (analysis object data) ROOT-based format into an easier-to-use MIT Open Data (MOD) HDF5 format~\cite{komiske_patrick_2019_3340205}.
Each SIM event consists of a list of particle flow candidates (PFCs), which are the reconstructed four-momentum and particle identification (PID) for each measured particle.
The PFCs are clustered into jets, using the anti-$k_t$ jet algorithm with $R = 0.5$~\cite{Cacciari:2005hq,Cacciari_2008,Cacciari:2011ma}.
For each jet, truth-level GEN jet information is also provided, as well as the CMS-prescribed JEC.
CMS-prescribed JERs are estimated using \Reference{Khachatryan_2017}.

We select jets whose GEN transverse momentum is in the range $p_T \in [500,1000]$ GeV.
The lower bound of $500$ GeV is to avoid any turn-on effects due to the $p_{T, \text{SIM}} > 375$ GeV cut applied to the dataset as a whole.
We require that the GEN pseudorapidity satisfies $|\eta| < 2.4$, and that jets are at least ``medium'' jet quality~\cite{CMS:2010xta}.
The latent variable of interest is $Z = p_{T,\text{GEN}}$, and the measured quantity $X = X_{\text{SIM}}$ depends on the choice of ML architecture.
All momenta are divided by a fixed scale of $1000$ GeV, so that the data values are roughly $\mathcal{O}(1)$.
In total, $5\times10^{6}$ jets are used for training, across the whole $p_T \in [500,1000]$ GeV range.

We consider four different ML models, of increasing sophistication:
\begin{enumerate}
    \item \textit{DNN}:
    The input $X$ consists only of the overall jet kinematic properties, with $X = (p_T, \eta, \phi)_\text{SIM}$, which is the same information used in the CMS calibration procedure in \Reference{Khachatryan_2017}.
    Each of the functions $A$, $B$, $C$, and $D$ are constructed as fully connected neural networks, with three hidden layers of size 64 and ReLU activations.
    \item \textit{EFN}:
    The input $X$ consists of the entire set of PFC three-momenta from the jet.
    Each of the functions $A$, $B$, $C$, and $D$ are constructed as Energy Flow Networks (EFNs)~\cite{Komiske_2019}.
    EFNs are permutation-invariant functions of point clouds, inspired by the Deep Sets formalism~\cite{NIPS2017_f22e4747}.
    They take the form $f(\{\Vec{p}_i\}) = F\left(\sum_i p_{Ti} \Phi(\eta_i,\phi_i)\right)$, which exhibits manifest infrared and colinear (IRC) safety.
    For each EFN, the $\Phi$ and $F$ functions consist of three hidden layers of respective sizes $(50,50,64)$ with ReLU activations.
    Since $C$ is a function of both $X$ and $Z$, the $Z$ is appended as an input to the $F$ function.
    \item \textit{PFN}: The same input features as the EFN, but inserted into a Particle Flow Network (PFN)~\cite{NIPS2017_f22e4747,Komiske_2019}, which does not impose IRC safety.
    PFNs take the form $f(\{\Vec{p}_i\}) = F\left(\sum_i  \Phi(p_{T_i}, \eta_i,\phi_i)\right)$. 
    \item \textit{PFN-PID}: The same as the PFN model, but in addition to the 3-momenta of each PFC, the reconstructed PID is included as an input feature.
    We follow the PID labeling scheme of \Reference{Komiske_2019} for photon, charged hadron, etc.
\end{enumerate}
Each of these models is trained for 200 epochs using the \Adam optimizer \cite{kingma2017adam}, with a learning rate of $\alpha = 10^{-4}$ and a batch size of 2048.
All model parameters are given an $L_2$ regularization loss with weight $\lambda_2 = 10^{-6}$.
The $D$ network is given an overall $L_1$ regularization loss of $\lambda_D = 10^{-3}$ to slowly force it to zero by the end of the training. 
Every 50 epochs, $\alpha$ is reduced by a factor of 5 and $\lambda_D$ is increased by a factor of 10.
To aid the numerical convergence, each model is pretrained with a mean squared error loss, $\mathcal{L}_{\text{pre}}[B,C] = \mathbb{E}_{P_{XZ}} \big[ (B(x)-z)^2 + (C(x,z) +  \text{cov}(X,Z))^2 \big]$.

\begin{table}[t]
    \centering
    \def\arraystretch{1.8}
    \begin{tabular}{r @{$\quad$} c @{$\quad$} c @{$\quad$} c @{$\quad$}   }
    \hline\hline
      Model  & Mean $\hat{p}_T$ [GeV] & Mean $\hat{\sigma}_{p_T}$ [GeV] & $I(X;Z)$ \\
      \hline
      DNN & $698 \pm 37.7$ & $35.7 \pm 2.1$ & 1.23 \\
      EFN  & $695 \pm 37.3$ & $32.6 \pm 2.3$ & 1.26 \\
      PFN  & $697 \pm 36.9$ & $32.5 \pm 2.5$ & 1.27 \\
      PFN-PID  & ${695 \pm 35.1}$ & $\mathbf{30.8 \pm 3.6}$ & \textbf{1.32} \\
      \hline
      CMS 2011 & $695 \pm 38.4$ & $36.9 \pm 1.7$ & --  \\

    \hline\hline
    \end{tabular}
    \caption{\GaussianAnsatz results for the four ML models, compared to the CMS 2011 baseline~\cite{Khachatryan_2017}.
    On a test dataset of GEN jets with $p_T \in [695,705]$ GeV, we show the inferred $\hat{p}_T$, its resolution $\hat{\sigma}_{p_T}$, and the learned mutual information between $X = X_{\text{SIM}}$ and $Z = p_{T,\text{GEN}}$.
    The $\pm$ values correspond to the standard deviation of the $\hat{p}_T$ and $\hat{\sigma}_{p_T}$ distributions across the test set, and bold face indicates the best resolution and highest mutual information.
    }
    \label{tab:JEC-results}
\end{table}

In \Tab{JEC-results}, we show the results of the training in a narrow bin of $p_{T, \text{GEN}} \in [695,705]$ GeV.
If our models yield unbiased estimators of the GEN $p_T$, then the inferred $\hat{p}_T$ distribution should be centered near 700 GeV, which it is for all models.
Adding more information to the model should not decrease the mutual information, and if useful, that information should improve the resolution.
We see indeed that the resolution improves with increasing model sophistication, as does the mutual information $I(X;Z)$.
The resolution from the DNN, which uses the same information as the CMS procedure, is marginally better than the nominal CMS 2011 jet resolution from \Reference{Khachatryan_2017}.
The PFN-PID model exhibits the best resolution, which is roughly 15\% better on average than the CMS baseline.

\begin{figure}[t]
    \centering
    \includegraphics[width=1.0\columnwidth]{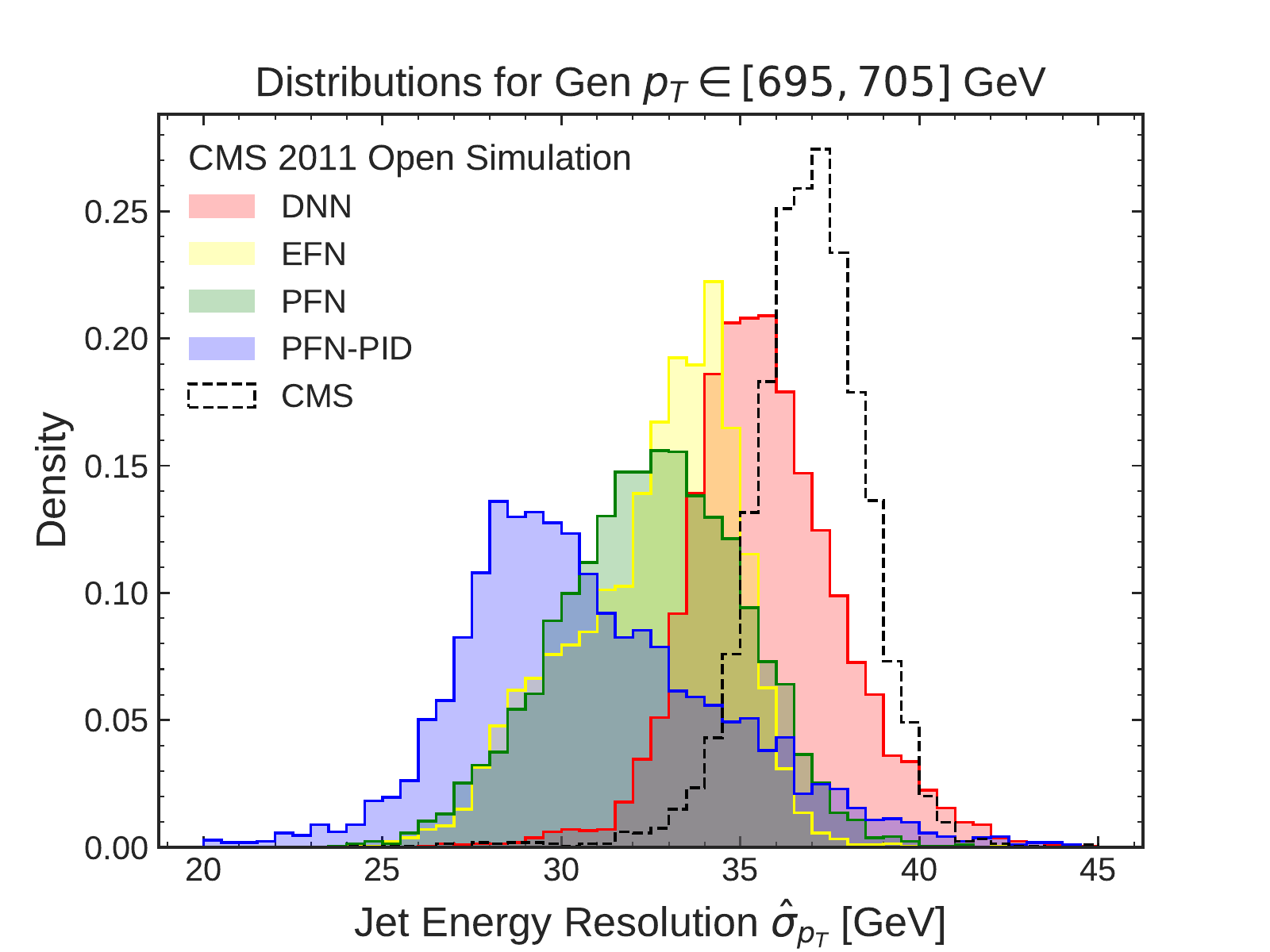}
    \caption{
    Learned JER distribution for the four models, compared to the CMS 2011 baseline.
    The dataset is the same as in \Tab{JEC-results}.
    On average, the PFN-PID exhibits 15\% better resolution (i.e.~smaller values) than the CMS default.}
    \label{fig:JEC-sigma}
\end{figure}

In \Fig{JEC-sigma}, we show the distribution of $\hat{\sigma}_{p_T}$ in the same $p_{T, \text{GEN}} \in [695,705]$ GeV bin.
As the model sophistication increases, the resolution increases (i.e.~the $\hat{\sigma}_{p_T}$ shift downward).
The non-Gaussian behavior of the ML models is expected, since these models are exploiting additional information beyond the $p_T$.
In principle, the resolution should never degrade by adding more information, but we do find a long right tail for the PFN-PID model due to incomplete ML convergence.%
\footnote{We verified that the tail shrinks and the resolution improves with increasing training statistics, but we were limited by machine memory considerations.}
We conclude that the measured PFC momenta, along with the PIDs, contain useful information for jet energy calibration that is lost when only considering the total jet momentum.

In this paper, we presented an extension of the MINE framework, the \GaussianAnsatz, capable of simultaneously
performing frequentist inference,
extracting Gaussian uncertainties,
and quantifying mutual information between random variables.
All of these tasks are performed in a single training, with no additional postprocessing.
Using this ML framework, we were able to take advantage of the full jet particle information in the CMS Open Simulation to improve the measured jet resolution by approximately $15\%$.
Studies by the ATLAS collaboration have used sequential calibration on a handful of observables to improve their resolution~\cite{ATLAS:2017bje,ATLAS:2014hvo,ATLAS:2019oxp}, and the Gaussian Ansatz may allow for further improvements by allowing for simultaneous calibrations of any number of input features.
We look forward to further developments in ML-based calibration and correlations methods in HEP and beyond.

\section*{Code and Data}

The code for the general-use \GaussianAnsatz framework can be found at \url{https://github.com/rikab/GaussianAnsatz}.
The code and data for the jet energy calibration study, in particular, are available at \url{https://github.com/rikab/GaussianAnsatz/tree/main/JEC}.

\section*{Acknowledgments}

We would like to thank Patrick Komiske for helpful discussions about EFNs and PFNs, Govert Nijs for helpful discussions on numerics and convergence, and Jennifer Roloff for helpful discussions about jet calibrations.
We are grateful to Phiala Shanahan and Andrew Pochinsky for providing access to the Wombat cluster for some of the calculations undertaken in this work.
RG and JT are supported by the National Science Foundation under Cooperative Agreement PHY-2019786 (The NSF AI Institute for Artificial Intelligence and Fundamental Interactions, \url{http://iaifi.org/}), and by the U.S. DOE Office of High Energy Physics under grant number DE-SC0012567.
BN is supported by the U.S. Department of Energy (DOE), Office of Science under contract DE-AC02-05CH11231.
\textbf{}
\bibliography{refs,HEPML}

\begin{thebibliography}{72}%
\makeatletter
\providecommand \@ifxundefined [1]{%
 \@ifx{#1\undefined}
}%
\providecommand \@ifnum [1]{%
 \ifnum #1\expandafter \@firstoftwo
 \else \expandafter \@secondoftwo
 \fi
}%
\providecommand \@ifx [1]{%
 \ifx #1\expandafter \@firstoftwo
 \else \expandafter \@secondoftwo
 \fi
}%
\providecommand \natexlab [1]{#1}%
\providecommand \enquote  [1]{``#1''}%
\providecommand \bibnamefont  [1]{#1}%
\providecommand \bibfnamefont [1]{#1}%
\providecommand \citenamefont [1]{#1}%
\providecommand \href@noop [0]{\@secondoftwo}%
\providecommand \href [0]{\begingroup \@sanitize@url \@href}%
\providecommand \@href[1]{\@@startlink{#1}\@@href}%
\providecommand \@@href[1]{\endgroup#1\@@endlink}%
\providecommand \@sanitize@url [0]{\catcode `\\12\catcode `\$12\catcode
  `\&12\catcode `\#12\catcode `\^12\catcode `\_12\catcode `\%12\relax}%
\providecommand \@@startlink[1]{}%
\providecommand \@@endlink[0]{}%
\providecommand \url  [0]{\begingroup\@sanitize@url \@url }%
\providecommand \@url [1]{\endgroup\@href {#1}{\urlprefix }}%
\providecommand \urlprefix  [0]{URL }%
\providecommand \Eprint [0]{\href }%
\providecommand \doibase [0]{http://dx.doi.org/}%
\providecommand \selectlanguage [0]{\@gobble}%
\providecommand \bibinfo  [0]{\@secondoftwo}%
\providecommand \bibfield  [0]{\@secondoftwo}%
\providecommand \translation [1]{[#1]}%
\providecommand \BibitemOpen [0]{}%
\providecommand \bibitemStop [0]{}%
\providecommand \bibitemNoStop [0]{.\EOS\space}%
\providecommand \EOS [0]{\spacefactor3000\relax}%
\providecommand \BibitemShut  [1]{\csname bibitem#1\endcsname}%
\let\auto@bib@innerbib\@empty
\bibitem [{\citenamefont {Sirunyan}\ \emph {et~al.}(2021)\citenamefont
  {Sirunyan} \emph {et~al.}}]{CMS:2020uim}%
  \BibitemOpen
  \bibfield  {author} {\bibinfo {author} {\bibfnamefont {Albert~M}\
  \bibnamefont {Sirunyan}} \emph {et~al.} (\bibinfo {collaboration} {CMS}),\
  }\bibfield  {title} {\enquote {\bibinfo {title} {{Electron and photon
  reconstruction and identification with the CMS experiment at the CERN
  LHC}},}\ }\href {\doibase 10.1088/1748-0221/16/05/P05014} {\bibfield
  {journal} {\bibinfo  {journal} {JINST}\ }\textbf {\bibinfo {volume} {16}},\
  \bibinfo {pages} {P05014} (\bibinfo {year} {2021})},\ \Eprint
  {http://arxiv.org/abs/2012.06888} {arXiv:2012.06888 [hep-ex]} \BibitemShut
  {NoStop}%
\bibitem [{\citenamefont {Kieseler}\ \emph {et~al.}(2021)\citenamefont
  {Kieseler}, \citenamefont {Strong}, \citenamefont {Chiandotto}, \citenamefont
  {Dorigo},\ and\ \citenamefont {Layer}}]{Kieseler:2021jxc}%
  \BibitemOpen
  \bibfield  {author} {\bibinfo {author} {\bibfnamefont {Jan}\ \bibnamefont
  {Kieseler}}, \bibinfo {author} {\bibfnamefont {Giles~C.}\ \bibnamefont
  {Strong}}, \bibinfo {author} {\bibfnamefont {Filippo}\ \bibnamefont
  {Chiandotto}}, \bibinfo {author} {\bibfnamefont {Tommaso}\ \bibnamefont
  {Dorigo}}, \ and\ \bibinfo {author} {\bibfnamefont {Lukas}\ \bibnamefont
  {Layer}},\ }\bibfield  {title} {\enquote {\bibinfo {title} {{Calorimetric
  Measurement of Multi-TeV Muons via Deep Regression}},}\ }\href@noop {} {\
  (\bibinfo {year} {2021})},\ \Eprint {http://arxiv.org/abs/2107.02119}
  {arXiv:2107.02119 [physics.ins-det]} \BibitemShut {NoStop}%
\bibitem [{\citenamefont {Belayneh}\ \emph {et~al.}(2020)\citenamefont
  {Belayneh} \emph {et~al.}}]{Belayneh:2019vyx}%
  \BibitemOpen
  \bibfield  {author} {\bibinfo {author} {\bibfnamefont {Dawit}\ \bibnamefont
  {Belayneh}} \emph {et~al.},\ }\bibfield  {title} {\enquote {\bibinfo {title}
  {{Calorimetry with deep learning: particle simulation and reconstruction for
  collider physics}},}\ }\href {\doibase 10.1140/epjc/s10052-020-8251-9}
  {\bibfield  {journal} {\bibinfo  {journal} {Eur. Phys. J. C}\ }\textbf
  {\bibinfo {volume} {80}},\ \bibinfo {pages} {688} (\bibinfo {year} {2020})},\
  \Eprint {http://arxiv.org/abs/1912.06794} {arXiv:1912.06794
  [physics.ins-det]} \BibitemShut {NoStop}%
\bibitem [{\citenamefont {{ATLAS
  Collaboration}}(2020{\natexlab{a}})}]{ATL-PHYS-PUB-2020-018}%
  \BibitemOpen
  \bibfield  {author} {\bibinfo {author} {\bibnamefont {{ATLAS
  Collaboration}}},\ }\bibfield  {title} {\enquote {\bibinfo {title} {{Deep
  Learning for Pion Identification and Energy Calibration with the ATLAS
  Detector}},}\ }\href {http://cdsweb.cern.ch/record/2724632} {\bibfield
  {journal} {\bibinfo  {journal} {ATL-PHYS-PUB-2020-018}\ } (\bibinfo {year}
  {2020}{\natexlab{a}})}\BibitemShut {NoStop}%
\bibitem [{\citenamefont {Akchurin}\ \emph
  {et~al.}(2021{\natexlab{a}})\citenamefont {Akchurin}, \citenamefont {Cowden},
  \citenamefont {Damgov}, \citenamefont {Hussain},\ and\ \citenamefont
  {Kunori}}]{Akchurin:2021afn}%
  \BibitemOpen
  \bibfield  {author} {\bibinfo {author} {\bibfnamefont {N.}~\bibnamefont
  {Akchurin}}, \bibinfo {author} {\bibfnamefont {C.}~\bibnamefont {Cowden}},
  \bibinfo {author} {\bibfnamefont {J.}~\bibnamefont {Damgov}}, \bibinfo
  {author} {\bibfnamefont {A.}~\bibnamefont {Hussain}}, \ and\ \bibinfo
  {author} {\bibfnamefont {S.}~\bibnamefont {Kunori}},\ }\bibfield  {title}
  {\enquote {\bibinfo {title} {{On the Use of Neural Networks for Energy
  Reconstruction in High-granularity Calorimeters}},}\ }\href@noop {} {\
  (\bibinfo {year} {2021}{\natexlab{a}})},\ \Eprint
  {http://arxiv.org/abs/2107.10207} {arXiv:2107.10207 [physics.ins-det]}
  \BibitemShut {NoStop}%
\bibitem [{\citenamefont {Akchurin}\ \emph
  {et~al.}(2021{\natexlab{b}})\citenamefont {Akchurin}, \citenamefont {Cowden},
  \citenamefont {Damgov}, \citenamefont {Hussain},\ and\ \citenamefont
  {Kunori}}]{Akchurin:2021ahx}%
  \BibitemOpen
  \bibfield  {author} {\bibinfo {author} {\bibfnamefont {N.}~\bibnamefont
  {Akchurin}}, \bibinfo {author} {\bibfnamefont {C.}~\bibnamefont {Cowden}},
  \bibinfo {author} {\bibfnamefont {J.}~\bibnamefont {Damgov}}, \bibinfo
  {author} {\bibfnamefont {A.}~\bibnamefont {Hussain}}, \ and\ \bibinfo
  {author} {\bibfnamefont {S.}~\bibnamefont {Kunori}},\ }\bibfield  {title}
  {\enquote {\bibinfo {title} {{Perspectives on the Calibration of CNN Energy
  Reconstruction in Highly Granular Calorimeters}},}\ }\href@noop {} {\
  (\bibinfo {year} {2021}{\natexlab{b}})},\ \Eprint
  {http://arxiv.org/abs/2108.10963} {arXiv:2108.10963 [physics.ins-det]}
  \BibitemShut {NoStop}%
\bibitem [{\citenamefont {Polson}\ \emph {et~al.}(2021)\citenamefont {Polson},
  \citenamefont {Kurchaninov},\ and\ \citenamefont
  {Lefebvre}}]{Polson:2021kvr}%
  \BibitemOpen
  \bibfield  {author} {\bibinfo {author} {\bibfnamefont {L.}~\bibnamefont
  {Polson}}, \bibinfo {author} {\bibfnamefont {L.}~\bibnamefont {Kurchaninov}},
  \ and\ \bibinfo {author} {\bibfnamefont {M.}~\bibnamefont {Lefebvre}},\
  }\bibfield  {title} {\enquote {\bibinfo {title} {{Energy reconstruction in a
  liquid argon calorimeter cell using convolutional neural networks}},}\
  }\href@noop {} {\  (\bibinfo {year} {2021})},\ \Eprint
  {http://arxiv.org/abs/2109.05124} {arXiv:2109.05124 [physics.ins-det]}
  \BibitemShut {NoStop}%
\bibitem [{\citenamefont {Pata}\ \emph {et~al.}(2021)\citenamefont {Pata},
  \citenamefont {Duarte}, \citenamefont {Vlimant}, \citenamefont {Pierini},\
  and\ \citenamefont {Spiropulu}}]{Pata:2021oez}%
  \BibitemOpen
  \bibfield  {author} {\bibinfo {author} {\bibfnamefont {Joosep}\ \bibnamefont
  {Pata}}, \bibinfo {author} {\bibfnamefont {Javier}\ \bibnamefont {Duarte}},
  \bibinfo {author} {\bibfnamefont {Jean-Roch}\ \bibnamefont {Vlimant}},
  \bibinfo {author} {\bibfnamefont {Maurizio}\ \bibnamefont {Pierini}}, \ and\
  \bibinfo {author} {\bibfnamefont {Maria}\ \bibnamefont {Spiropulu}},\
  }\bibfield  {title} {\enquote {\bibinfo {title} {{MLPF: Efficient
  machine-learned particle-flow reconstruction using graph neural networks}},}\
  }\href@noop {} {\  (\bibinfo {year} {2021})},\ \Eprint
  {http://arxiv.org/abs/2101.08578} {arXiv:2101.08578 [physics.data-an]}
  \BibitemShut {NoStop}%
\bibitem [{\citenamefont {{ATLAS
  Collaboration}}(2018)}]{ATL-PHYS-PUB-2018-013}%
  \BibitemOpen
  \bibfield  {author} {\bibinfo {author} {\bibnamefont {{ATLAS
  Collaboration}}},\ }\bibfield  {title} {\enquote {\bibinfo {title}
  {{Generalized Numerical Inversion: A Neural Network Approach to Jet
  Calibration}},}\ }\href {http://cdsweb.cern.ch/record/2630972} {\bibfield
  {journal} {\bibinfo  {journal} {ATL-PHYS-PUB-2018-013}\ } (\bibinfo {year}
  {2018})}\BibitemShut {NoStop}%
\bibitem [{\citenamefont {{ATLAS
  Collaboration}}(2020{\natexlab{b}})}]{ATL-PHYS-PUB-2020-001}%
  \BibitemOpen
  \bibfield  {author} {\bibinfo {author} {\bibnamefont {{ATLAS
  Collaboration}}},\ }\bibfield  {title} {\enquote {\bibinfo {title}
  {{Simultaneous Jet Energy and Mass Calibrations with Neural Networks}},}\
  }\href {http://cdsweb.cern.ch/record/2706189} {\bibfield  {journal} {\bibinfo
   {journal} {ATL-PHYS-PUB-2020-001}\ } (\bibinfo {year}
  {2020}{\natexlab{b}})}\BibitemShut {NoStop}%
\bibitem [{\citenamefont {Sirunyan}\ \emph {et~al.}(2020)\citenamefont
  {Sirunyan} \emph {et~al.}}]{CMS:2019uxx}%
  \BibitemOpen
  \bibfield  {author} {\bibinfo {author} {\bibfnamefont {Albert~M}\
  \bibnamefont {Sirunyan}} \emph {et~al.} (\bibinfo {collaboration} {CMS}),\
  }\bibfield  {title} {\enquote {\bibinfo {title} {{A Deep Neural Network for
  Simultaneous Estimation of b Jet Energy and Resolution}},}\ }\href {\doibase
  10.1007/s41781-020-00041-z} {\bibfield  {journal} {\bibinfo  {journal}
  {Comput. Softw. Big Sci.}\ }\textbf {\bibinfo {volume} {4}},\ \bibinfo
  {pages} {10} (\bibinfo {year} {2020})},\ \Eprint
  {http://arxiv.org/abs/1912.06046} {arXiv:1912.06046 [hep-ex]} \BibitemShut
  {NoStop}%
\bibitem [{\citenamefont {Haake}\ and\ \citenamefont
  {Loizides}(2019)}]{Haake:2018hqn}%
  \BibitemOpen
  \bibfield  {author} {\bibinfo {author} {\bibfnamefont {R\"udiger}\
  \bibnamefont {Haake}}\ and\ \bibinfo {author} {\bibfnamefont {Constantin}\
  \bibnamefont {Loizides}},\ }\bibfield  {title} {\enquote {\bibinfo {title}
  {{Machine Learning based jet momentum reconstruction in heavy-ion
  collisions}},}\ }\href {\doibase 10.1103/PhysRevC.99.064904} {\bibfield
  {journal} {\bibinfo  {journal} {Phys. Rev. C}\ }\textbf {\bibinfo {volume}
  {99}},\ \bibinfo {pages} {064904} (\bibinfo {year} {2019})},\ \Eprint
  {http://arxiv.org/abs/1810.06324} {arXiv:1810.06324 [nucl-ex]} \BibitemShut
  {NoStop}%
\bibitem [{\citenamefont {Haake}(2020)}]{Haake:2019pqd}%
  \BibitemOpen
  \bibfield  {author} {\bibinfo {author} {\bibfnamefont {R\"udiger}\
  \bibnamefont {Haake}} (\bibinfo {collaboration} {ALICE}),\ }\bibfield
  {title} {\enquote {\bibinfo {title} {{Machine Learning based jet momentum
  reconstruction in Pb-Pb collisions measured with the ALICE detector}},}\
  }\href {\doibase 10.22323/1.364.0312} {\bibfield  {journal} {\bibinfo
  {journal} {PoS}\ }\textbf {\bibinfo {volume} {EPS-HEP2019}},\ \bibinfo
  {pages} {312} (\bibinfo {year} {2020})},\ \Eprint
  {http://arxiv.org/abs/1909.01639} {arXiv:1909.01639 [nucl-ex]} \BibitemShut
  {NoStop}%
\bibitem [{\citenamefont {Baldi}\ \emph {et~al.}(2020)\citenamefont {Baldi},
  \citenamefont {Blecher}, \citenamefont {Butter}, \citenamefont {Collado},
  \citenamefont {Howard}, \citenamefont {Keilbach}, \citenamefont {Plehn},
  \citenamefont {Kasieczka},\ and\ \citenamefont {Whiteson}}]{Baldi:2020hjm}%
  \BibitemOpen
  \bibfield  {author} {\bibinfo {author} {\bibfnamefont {Pierre}\ \bibnamefont
  {Baldi}}, \bibinfo {author} {\bibfnamefont {Lukas}\ \bibnamefont {Blecher}},
  \bibinfo {author} {\bibfnamefont {Anja}\ \bibnamefont {Butter}}, \bibinfo
  {author} {\bibfnamefont {Julian}\ \bibnamefont {Collado}}, \bibinfo {author}
  {\bibfnamefont {Jessica~N.}\ \bibnamefont {Howard}}, \bibinfo {author}
  {\bibfnamefont {Fabian}\ \bibnamefont {Keilbach}}, \bibinfo {author}
  {\bibfnamefont {Tilman}\ \bibnamefont {Plehn}}, \bibinfo {author}
  {\bibfnamefont {Gregor}\ \bibnamefont {Kasieczka}}, \ and\ \bibinfo {author}
  {\bibfnamefont {Daniel}\ \bibnamefont {Whiteson}},\ }\bibfield  {title}
  {\enquote {\bibinfo {title} {{How to GAN Higher Jet Resolution}},}\
  }\href@noop {} {\  (\bibinfo {year} {2020})},\ \Eprint
  {http://arxiv.org/abs/2012.11944} {arXiv:2012.11944 [hep-ph]} \BibitemShut
  {NoStop}%
\bibitem [{\citenamefont {Komiske}\ \emph {et~al.}(2017)\citenamefont
  {Komiske}, \citenamefont {Metodiev}, \citenamefont {Nachman},\ and\
  \citenamefont {Schwartz}}]{Komiske:2017ubm}%
  \BibitemOpen
  \bibfield  {author} {\bibinfo {author} {\bibfnamefont {Patrick~T.}\
  \bibnamefont {Komiske}}, \bibinfo {author} {\bibfnamefont {Eric~M.}\
  \bibnamefont {Metodiev}}, \bibinfo {author} {\bibfnamefont {Benjamin}\
  \bibnamefont {Nachman}}, \ and\ \bibinfo {author} {\bibfnamefont
  {Matthew~D.}\ \bibnamefont {Schwartz}},\ }\bibfield  {title} {\enquote
  {\bibinfo {title} {{Pileup Mitigation with Machine Learning (PUMML)}},}\
  }\href {\doibase 10.1007/JHEP12(2017)051} {\bibfield  {journal} {\bibinfo
  {journal} {JHEP}\ }\textbf {\bibinfo {volume} {12}},\ \bibinfo {pages} {051}
  (\bibinfo {year} {2017})},\ \Eprint {http://arxiv.org/abs/1707.08600}
  {arXiv:1707.08600 [hep-ph]} \BibitemShut {NoStop}%
\bibitem [{ATL(2019)}]{ATL-PHYS-PUB-2019-028}%
  \BibitemOpen
  \href {https://cds.cern.ch/record/2684070} {\emph {\bibinfo {title}
  {{Convolutional Neural Networks with Event Images for Pileup Mitigation with
  the ATLAS Detector}}}},\ \bibinfo {type} {Tech. Rep.}\ (\bibinfo
  {institution} {CERN},\ \bibinfo {address} {Geneva},\ \bibinfo {year}
  {2019})\BibitemShut {NoStop}%
\bibitem [{\citenamefont {Maier}\ \emph {et~al.}(2021)\citenamefont {Maier},
  \citenamefont {Narayanan}, \citenamefont {de~Castro}, \citenamefont
  {Goncharov}, \citenamefont {Paus},\ and\ \citenamefont
  {Schott}}]{Maier:2021ymx}%
  \BibitemOpen
  \bibfield  {author} {\bibinfo {author} {\bibfnamefont {Benedikt}\
  \bibnamefont {Maier}}, \bibinfo {author} {\bibfnamefont {Siddharth~M.}\
  \bibnamefont {Narayanan}}, \bibinfo {author} {\bibfnamefont {Gianfranco}\
  \bibnamefont {de~Castro}}, \bibinfo {author} {\bibfnamefont {Maxim}\
  \bibnamefont {Goncharov}}, \bibinfo {author} {\bibfnamefont {Christoph}\
  \bibnamefont {Paus}}, \ and\ \bibinfo {author} {\bibfnamefont {Matthias}\
  \bibnamefont {Schott}},\ }\bibfield  {title} {\enquote {\bibinfo {title}
  {{Pile-Up Mitigation using Attention}},}\ }\href@noop {} {\  (\bibinfo {year}
  {2021})},\ \Eprint {http://arxiv.org/abs/2107.02779} {arXiv:2107.02779
  [physics.ins-det]} \BibitemShut {NoStop}%
\bibitem [{\citenamefont {Kasieczka}\ \emph {et~al.}(2020)\citenamefont
  {Kasieczka}, \citenamefont {Luchmann}, \citenamefont {Otterpohl},\ and\
  \citenamefont {Plehn}}]{Kasieczka:2020vlh}%
  \BibitemOpen
  \bibfield  {author} {\bibinfo {author} {\bibfnamefont {Gregor}\ \bibnamefont
  {Kasieczka}}, \bibinfo {author} {\bibfnamefont {Michel}\ \bibnamefont
  {Luchmann}}, \bibinfo {author} {\bibfnamefont {Florian}\ \bibnamefont
  {Otterpohl}}, \ and\ \bibinfo {author} {\bibfnamefont {Tilman}\ \bibnamefont
  {Plehn}},\ }\bibfield  {title} {\enquote {\bibinfo {title} {{Per-Object
  Systematics using Deep-Learned Calibration}},}\ }\href {\doibase
  10.21468/SciPostPhys.9.6.089} {\  (\bibinfo {year} {2020}),\
  10.21468/SciPostPhys.9.6.089},\ \Eprint {http://arxiv.org/abs/2003.11099}
  {arXiv:2003.11099 [hep-ph]} \BibitemShut {NoStop}%
\bibitem [{\citenamefont {Arjona~Mart\'\i{}nez}\ \emph
  {et~al.}(2019)\citenamefont {Arjona~Mart\'\i{}nez}, \citenamefont {Cerri},
  \citenamefont {Pierini}, \citenamefont {Spiropulu},\ and\ \citenamefont
  {Vlimant}}]{ArjonaMartinez:2018eah}%
  \BibitemOpen
  \bibfield  {author} {\bibinfo {author} {\bibfnamefont {J.}~\bibnamefont
  {Arjona~Mart\'\i{}nez}}, \bibinfo {author} {\bibfnamefont {Olmo}\
  \bibnamefont {Cerri}}, \bibinfo {author} {\bibfnamefont {Maurizio}\
  \bibnamefont {Pierini}}, \bibinfo {author} {\bibfnamefont {Maria}\
  \bibnamefont {Spiropulu}}, \ and\ \bibinfo {author} {\bibfnamefont
  {Jean-Roch}\ \bibnamefont {Vlimant}},\ }\bibfield  {title} {\enquote
  {\bibinfo {title} {{Pileup mitigation at the Large Hadron Collider with graph
  neural networks}},}\ }\href {\doibase 10.1140/epjp/i2019-12710-3} {\bibfield
  {journal} {\bibinfo  {journal} {Eur. Phys. J. Plus}\ }\textbf {\bibinfo
  {volume} {134}},\ \bibinfo {pages} {333} (\bibinfo {year} {2019})},\ \Eprint
  {http://arxiv.org/abs/1810.07988} {arXiv:1810.07988 [hep-ph]} \BibitemShut
  {NoStop}%
\bibitem [{\citenamefont {Holmberg}(2022)}]{jec_with_gnn_regression}%
  \BibitemOpen
  \bibfield  {author} {\bibinfo {author} {\bibfnamefont {Daniel}\ \bibnamefont
  {Holmberg}},\ }\emph {\bibinfo {title} {Jet Energy Corrections with Graph
  Neural Network Regression}},\ \href@noop {} {Master's thesis},\ \bibinfo
  {school} {University of Helsinki} (\bibinfo {year} {2022})\BibitemShut
  {NoStop}%
\bibitem [{\citenamefont {Diefenthaler}\ \emph {et~al.}(2021)\citenamefont
  {Diefenthaler}, \citenamefont {Farhat}, \citenamefont {Verbytskyi},\ and\
  \citenamefont {Xu}}]{Diefenthaler:2021rdj}%
  \BibitemOpen
  \bibfield  {author} {\bibinfo {author} {\bibfnamefont {Markus}\ \bibnamefont
  {Diefenthaler}}, \bibinfo {author} {\bibfnamefont {Abduhhal}\ \bibnamefont
  {Farhat}}, \bibinfo {author} {\bibfnamefont {Andrii}\ \bibnamefont
  {Verbytskyi}}, \ and\ \bibinfo {author} {\bibfnamefont {Yuesheng}\
  \bibnamefont {Xu}},\ }\bibfield  {title} {\enquote {\bibinfo {title} {{Deeply
  Learning Deep Inelastic Scattering Kinematics}},}\ }\href@noop {} {\
  (\bibinfo {year} {2021})},\ \Eprint {http://arxiv.org/abs/2108.11638}
  {arXiv:2108.11638 [hep-ph]} \BibitemShut {NoStop}%
\bibitem [{\citenamefont {Arratia}\ \emph {et~al.}(2021)\citenamefont
  {Arratia}, \citenamefont {Britzger}, \citenamefont {Long},\ and\
  \citenamefont {Nachman}}]{Arratia:2021tsq}%
  \BibitemOpen
  \bibfield  {author} {\bibinfo {author} {\bibfnamefont {Miguel}\ \bibnamefont
  {Arratia}}, \bibinfo {author} {\bibfnamefont {Daniel}\ \bibnamefont
  {Britzger}}, \bibinfo {author} {\bibfnamefont {Owen}\ \bibnamefont {Long}}, \
  and\ \bibinfo {author} {\bibfnamefont {Benjamin}\ \bibnamefont {Nachman}},\
  }\bibfield  {title} {\enquote {\bibinfo {title} {{Reconstructing the
  Kinematics of Deep Inelastic Scattering with Deep Learning}},}\ }\href@noop
  {} {\  (\bibinfo {year} {2021})},\ \Eprint {http://arxiv.org/abs/2110.05505}
  {arXiv:2110.05505 [hep-ex]} \BibitemShut {NoStop}%
\bibitem [{\citenamefont {Liu}\ \emph {et~al.}(2020)\citenamefont {Liu},
  \citenamefont {Ott}, \citenamefont {Collado}, \citenamefont {Jargowsky},
  \citenamefont {Wu}, \citenamefont {Bian},\ and\ \citenamefont
  {Baldi}}]{Liu:2020pzv}%
  \BibitemOpen
  \bibfield  {author} {\bibinfo {author} {\bibfnamefont {Junze}\ \bibnamefont
  {Liu}}, \bibinfo {author} {\bibfnamefont {Jordan}\ \bibnamefont {Ott}},
  \bibinfo {author} {\bibfnamefont {Julian}\ \bibnamefont {Collado}}, \bibinfo
  {author} {\bibfnamefont {Benjamin}\ \bibnamefont {Jargowsky}}, \bibinfo
  {author} {\bibfnamefont {Wenjie}\ \bibnamefont {Wu}}, \bibinfo {author}
  {\bibfnamefont {Jianming}\ \bibnamefont {Bian}}, \ and\ \bibinfo {author}
  {\bibfnamefont {Pierre}\ \bibnamefont {Baldi}} (\bibinfo {collaboration}
  {DUNE}),\ }\bibfield  {title} {\enquote {\bibinfo {title}
  {{Deep-Learning-Based Kinematic Reconstruction for DUNE}},}\ }\href@noop {}
  {\  (\bibinfo {year} {2020})},\ \Eprint {http://arxiv.org/abs/2012.06181}
  {arXiv:2012.06181 [physics.ins-det]} \BibitemShut {NoStop}%
\bibitem [{\citenamefont {Delaquis}\ \emph {et~al.}(2018)\citenamefont
  {Delaquis} \emph {et~al.}}]{EXO:2018bpx}%
  \BibitemOpen
  \bibfield  {author} {\bibinfo {author} {\bibfnamefont {S.}~\bibnamefont
  {Delaquis}} \emph {et~al.} (\bibinfo {collaboration} {EXO}),\ }\bibfield
  {title} {\enquote {\bibinfo {title} {{Deep Neural Networks for Energy and
  Position Reconstruction in EXO-200}},}\ }\href {\doibase
  10.1088/1748-0221/13/08/P08023} {\bibfield  {journal} {\bibinfo  {journal}
  {JINST}\ }\textbf {\bibinfo {volume} {13}},\ \bibinfo {pages} {P08023}
  (\bibinfo {year} {2018})},\ \Eprint {http://arxiv.org/abs/1804.09641}
  {arXiv:1804.09641 [physics.ins-det]} \BibitemShut {NoStop}%
\bibitem [{\citenamefont {Baldi}\ \emph {et~al.}(2019)\citenamefont {Baldi},
  \citenamefont {Bian}, \citenamefont {Hertel},\ and\ \citenamefont
  {Li}}]{Baldi:2018qhe}%
  \BibitemOpen
  \bibfield  {author} {\bibinfo {author} {\bibfnamefont {Pierre}\ \bibnamefont
  {Baldi}}, \bibinfo {author} {\bibfnamefont {Jianming}\ \bibnamefont {Bian}},
  \bibinfo {author} {\bibfnamefont {Lars}\ \bibnamefont {Hertel}}, \ and\
  \bibinfo {author} {\bibfnamefont {Lingge}\ \bibnamefont {Li}},\ }\bibfield
  {title} {\enquote {\bibinfo {title} {{Improved Energy Reconstruction in NOvA
  with Regression Convolutional Neural Networks}},}\ }\href {\doibase
  10.1103/PhysRevD.99.012011} {\bibfield  {journal} {\bibinfo  {journal} {Phys.
  Rev. D}\ }\textbf {\bibinfo {volume} {99}},\ \bibinfo {pages} {012011}
  (\bibinfo {year} {2019})},\ \Eprint {http://arxiv.org/abs/1811.04557}
  {arXiv:1811.04557 [physics.ins-det]} \BibitemShut {NoStop}%
\bibitem [{\citenamefont {Abbasi}\ \emph {et~al.}()\citenamefont {Abbasi} \emph
  {et~al.}}]{Abbasi:2021ryj}%
  \BibitemOpen
  \bibfield  {author} {\bibinfo {author} {\bibfnamefont {R.}~\bibnamefont
  {Abbasi}} \emph {et~al.},\ }\bibfield  {title} {\enquote {\bibinfo {title}
  {{A Convolutional Neural Network based Cascade Reconstruction for the IceCube
  Neutrino Observatory}},}\ }\href {\doibase 10.1088/1748-0221/16/07/P07041}
  {\bibfield  {journal} {\bibinfo  {journal} {JINST}\ }\textbf {\bibinfo
  {volume} {16}},\ \bibinfo {pages} {P07041}},\ \Eprint
  {http://arxiv.org/abs/2101.11589} {arXiv:2101.11589 [hep-ex]} \BibitemShut
  {NoStop}%
\bibitem [{\citenamefont {Aartsen}\ \emph {et~al.}(2020)\citenamefont {Aartsen}
  \emph {et~al.}}]{IceCube:2020yct}%
  \BibitemOpen
  \bibfield  {author} {\bibinfo {author} {\bibfnamefont {M.~G.}\ \bibnamefont
  {Aartsen}} \emph {et~al.} (\bibinfo {collaboration} {IceCube}),\ }\bibfield
  {title} {\enquote {\bibinfo {title} {{Cosmic ray spectrum from 250 TeV to 10
  PeV using IceTop}},}\ }\href {\doibase 10.1103/PhysRevD.102.122001}
  {\bibfield  {journal} {\bibinfo  {journal} {Phys. Rev. D}\ }\textbf {\bibinfo
  {volume} {102}},\ \bibinfo {pages} {122001} (\bibinfo {year} {2020})},\
  \Eprint {http://arxiv.org/abs/2006.05215} {arXiv:2006.05215 [astro-ph.HE]}
  \BibitemShut {NoStop}%
\bibitem [{\citenamefont {Carloni}\ \emph {et~al.}(2021)\citenamefont
  {Carloni}, \citenamefont {Kamp}, \citenamefont {Schneider},\ and\
  \citenamefont {Conrad}}]{Carloni:2021zbc}%
  \BibitemOpen
  \bibfield  {author} {\bibinfo {author} {\bibfnamefont {Kiara}\ \bibnamefont
  {Carloni}}, \bibinfo {author} {\bibfnamefont {Nicholas~W.}\ \bibnamefont
  {Kamp}}, \bibinfo {author} {\bibfnamefont {Austin}\ \bibnamefont
  {Schneider}}, \ and\ \bibinfo {author} {\bibfnamefont {Janet~M.}\
  \bibnamefont {Conrad}},\ }\bibfield  {title} {\enquote {\bibinfo {title}
  {{Convolutional Neural Networks for Shower Energy Prediction in Liquid Argon
  Time Projection Chambers}},}\ }\href@noop {} {\  (\bibinfo {year} {2021})},\
  \Eprint {http://arxiv.org/abs/2110.10766} {arXiv:2110.10766 [hep-ex]}
  \BibitemShut {NoStop}%
\bibitem [{\citenamefont {Feickert}\ and\ \citenamefont
  {Nachman}(2021)}]{Feickert:2021ajf}%
  \BibitemOpen
  \bibfield  {author} {\bibinfo {author} {\bibfnamefont {Matthew}\ \bibnamefont
  {Feickert}}\ and\ \bibinfo {author} {\bibfnamefont {Benjamin}\ \bibnamefont
  {Nachman}},\ }\bibfield  {title} {\enquote {\bibinfo {title} {{A Living
  Review of Machine Learning for Particle Physics}},}\ }\href@noop {} {\
  (\bibinfo {year} {2021})},\ \Eprint {http://arxiv.org/abs/2102.02770}
  {arXiv:2102.02770 [hep-ph]} \BibitemShut {NoStop}%
\bibitem [{\citenamefont {Gambhir}\ \emph {et~al.}(2022)\citenamefont
  {Gambhir}, \citenamefont {Nachman},\ and\ \citenamefont
  {Thaler}}]{priordependence}%
  \BibitemOpen
  \bibfield  {author} {\bibinfo {author} {\bibfnamefont {Rikab}\ \bibnamefont
  {Gambhir}}, \bibinfo {author} {\bibfnamefont {Benjamin}\ \bibnamefont
  {Nachman}}, \ and\ \bibinfo {author} {\bibfnamefont {Jesse}\ \bibnamefont
  {Thaler}},\ }\href {\doibase 10.48550/ARXIV.2205.05084} {\enquote {\bibinfo
  {title} {Bias and priors in machine learning calibrations for high energy
  physics},}\ } (\bibinfo {year} {2022}),\ \Eprint
  {http://arxiv.org/abs/2205.05084} {arXiv:2205.05084 [hep-ph]} \BibitemShut
  {NoStop}%
\bibitem [{\citenamefont {Sirunyan}\ \emph
  {et~al.}(2019{\natexlab{a}})\citenamefont {Sirunyan} \emph
  {et~al.}}]{CMS:2019ctu}%
  \BibitemOpen
  \bibfield  {author} {\bibinfo {author} {\bibfnamefont {Albert~M}\
  \bibnamefont {Sirunyan}} \emph {et~al.} (\bibinfo {collaboration} {CMS}),\
  }\bibfield  {title} {\enquote {\bibinfo {title} {{Performance of missing
  transverse momentum reconstruction in proton-proton collisions at $\sqrt{s}
  =$ 13 TeV using the CMS detector}},}\ }\href {\doibase
  10.1088/1748-0221/14/07/P07004} {\bibfield  {journal} {\bibinfo  {journal}
  {JINST}\ }\textbf {\bibinfo {volume} {14}},\ \bibinfo {pages} {P07004}
  (\bibinfo {year} {2019}{\natexlab{a}})},\ \Eprint
  {http://arxiv.org/abs/1903.06078} {arXiv:1903.06078 [hep-ex]} \BibitemShut
  {NoStop}%
\bibitem [{\citenamefont {Nachman}\ and\ \citenamefont
  {Lester}(2013)}]{Nachman:2013bia}%
  \BibitemOpen
  \bibfield  {author} {\bibinfo {author} {\bibfnamefont {Benjamin}\
  \bibnamefont {Nachman}}\ and\ \bibinfo {author} {\bibfnamefont
  {Christopher~G.}\ \bibnamefont {Lester}},\ }\bibfield  {title} {\enquote
  {\bibinfo {title} {{Significance Variables}},}\ }\href {\doibase
  10.1103/PhysRevD.88.075013} {\bibfield  {journal} {\bibinfo  {journal} {Phys.
  Rev. D}\ }\textbf {\bibinfo {volume} {88}},\ \bibinfo {pages} {075013}
  (\bibinfo {year} {2013})},\ \Eprint {http://arxiv.org/abs/1303.7009}
  {arXiv:1303.7009 [hep-ph]} \BibitemShut {NoStop}%
\bibitem [{\citenamefont {Aad}\ \emph {et~al.}(2013)\citenamefont {Aad} \emph
  {et~al.}}]{ATLAS:2012qgw}%
  \BibitemOpen
  \bibfield  {author} {\bibinfo {author} {\bibfnamefont {Georges}\ \bibnamefont
  {Aad}} \emph {et~al.} (\bibinfo {collaboration} {ATLAS}),\ }\bibfield
  {title} {\enquote {\bibinfo {title} {{Search for squarks and gluinos with the
  ATLAS detector in final states with jets and missing transverse momentum
  using 4.7 fb$^{-1}$ of $\sqrt{s}=7$ TeV proton-proton collision data}},}\
  }\href {\doibase 10.1103/PhysRevD.87.012008} {\bibfield  {journal} {\bibinfo
  {journal} {Phys. Rev. D}\ }\textbf {\bibinfo {volume} {87}},\ \bibinfo
  {pages} {012008} (\bibinfo {year} {2013})},\ \Eprint
  {http://arxiv.org/abs/1208.0949} {arXiv:1208.0949 [hep-ex]} \BibitemShut
  {NoStop}%
\bibitem [{\citenamefont {Aad}\ \emph {et~al.}(2021)\citenamefont {Aad} \emph
  {et~al.}}]{ATLAS:2021kxv}%
  \BibitemOpen
  \bibfield  {author} {\bibinfo {author} {\bibfnamefont {Georges}\ \bibnamefont
  {Aad}} \emph {et~al.} (\bibinfo {collaboration} {ATLAS}),\ }\bibfield
  {title} {\enquote {\bibinfo {title} {{Search for new phenomena in events with
  an energetic jet and missing transverse momentum in $pp$ collisions at $\sqrt
  {s}$ =13 TeV with the ATLAS detector}},}\ }\href {\doibase
  10.1103/PhysRevD.103.112006} {\bibfield  {journal} {\bibinfo  {journal}
  {Phys. Rev. D}\ }\textbf {\bibinfo {volume} {103}},\ \bibinfo {pages}
  {112006} (\bibinfo {year} {2021})},\ \Eprint
  {http://arxiv.org/abs/2102.10874} {arXiv:2102.10874 [hep-ex]} \BibitemShut
  {NoStop}%
\bibitem [{\citenamefont {Sirunyan}\ \emph
  {et~al.}(2019{\natexlab{b}})\citenamefont {Sirunyan} \emph
  {et~al.}}]{Sirunyan:2019wwa}%
  \BibitemOpen
  \bibfield  {author} {\bibinfo {author} {\bibfnamefont {Albert~M}\
  \bibnamefont {Sirunyan}} \emph {et~al.} (\bibinfo {collaboration} {CMS}),\
  }\bibfield  {title} {\enquote {\bibinfo {title} {{A deep neural network for
  simultaneous estimation of b jet energy and resolution}},}\ }\href {\doibase
  10.1007/s41781-020-00041-z} {\  (\bibinfo {year} {2019}{\natexlab{b}}),\
  10.1007/s41781-020-00041-z},\ \Eprint {http://arxiv.org/abs/1912.06046}
  {arXiv:1912.06046 [hep-ex]} \BibitemShut {NoStop}%
\bibitem [{\citenamefont {Cheong}\ \emph {et~al.}(2020)\citenamefont {Cheong},
  \citenamefont {Cukierman}, \citenamefont {Nachman}, \citenamefont {Safdari},\
  and\ \citenamefont {Schwartzman}}]{Cheong:2019upg}%
  \BibitemOpen
  \bibfield  {author} {\bibinfo {author} {\bibfnamefont {Sanha}\ \bibnamefont
  {Cheong}}, \bibinfo {author} {\bibfnamefont {Aviv}\ \bibnamefont
  {Cukierman}}, \bibinfo {author} {\bibfnamefont {Benjamin}\ \bibnamefont
  {Nachman}}, \bibinfo {author} {\bibfnamefont {Murtaza}\ \bibnamefont
  {Safdari}}, \ and\ \bibinfo {author} {\bibfnamefont {Ariel}\ \bibnamefont
  {Schwartzman}},\ }\bibfield  {title} {\enquote {\bibinfo {title}
  {{Parametrizing the Detector Response with Neural Networks}},}\ }\href
  {\doibase 10.1088/1748-0221/15/01/P01030} {\bibfield  {journal} {\bibinfo
  {journal} {JINST}\ }\textbf {\bibinfo {volume} {15}},\ \bibinfo {pages}
  {P01030} (\bibinfo {year} {2020})},\ \Eprint
  {http://arxiv.org/abs/1910.03773} {arXiv:1910.03773 [physics.data-an]}
  \BibitemShut {NoStop}%
\bibitem [{\citenamefont {Bollweg}\ \emph {et~al.}(2020)\citenamefont
  {Bollweg}, \citenamefont {Haußmann}, \citenamefont {Kasieczka},
  \citenamefont {Luchmann}, \citenamefont {Plehn},\ and\ \citenamefont
  {Thompson}}]{Bollweg:2019skg}%
  \BibitemOpen
  \bibfield  {author} {\bibinfo {author} {\bibfnamefont {Sven}\ \bibnamefont
  {Bollweg}}, \bibinfo {author} {\bibfnamefont {Manuel}\ \bibnamefont
  {Haußmann}}, \bibinfo {author} {\bibfnamefont {Gregor}\ \bibnamefont
  {Kasieczka}}, \bibinfo {author} {\bibfnamefont {Michel}\ \bibnamefont
  {Luchmann}}, \bibinfo {author} {\bibfnamefont {Tilman}\ \bibnamefont
  {Plehn}}, \ and\ \bibinfo {author} {\bibfnamefont {Jennifer}\ \bibnamefont
  {Thompson}},\ }\bibfield  {title} {\enquote {\bibinfo {title} {{Deep-Learning
  Jets with Uncertainties and More}},}\ }\href {\doibase
  10.21468/SciPostPhys.8.1.006} {\bibfield  {journal} {\bibinfo  {journal}
  {SciPost Phys.}\ }\textbf {\bibinfo {volume} {8}},\ \bibinfo {pages} {006}
  (\bibinfo {year} {2020})},\ \Eprint {http://arxiv.org/abs/1904.10004}
  {arXiv:1904.10004 [hep-ph]} \BibitemShut {NoStop}%
\bibitem [{\citenamefont {Bellagente}\ \emph {et~al.}(2021)\citenamefont
  {Bellagente}, \citenamefont {Hau\ss{}mann}, \citenamefont {Luchmann},\ and\
  \citenamefont {Plehn}}]{Bellagente:2021yyh}%
  \BibitemOpen
  \bibfield  {author} {\bibinfo {author} {\bibfnamefont {Marco}\ \bibnamefont
  {Bellagente}}, \bibinfo {author} {\bibfnamefont {Manuel}\ \bibnamefont
  {Hau\ss{}mann}}, \bibinfo {author} {\bibfnamefont {Michel}\ \bibnamefont
  {Luchmann}}, \ and\ \bibinfo {author} {\bibfnamefont {Tilman}\ \bibnamefont
  {Plehn}},\ }\bibfield  {title} {\enquote {\bibinfo {title} {{Understanding
  Event-Generation Networks via Uncertainties}},}\ }\href@noop {} {\  (\bibinfo
  {year} {2021})},\ \Eprint {http://arxiv.org/abs/2104.04543} {arXiv:2104.04543
  [hep-ph]} \BibitemShut {NoStop}%
\bibitem [{\citenamefont {Kronheim}\ \emph {et~al.}(2020)\citenamefont
  {Kronheim}, \citenamefont {Kuchera}, \citenamefont {Prosper},\ and\
  \citenamefont {Karbo}}]{1806026}%
  \BibitemOpen
  \bibfield  {author} {\bibinfo {author} {\bibfnamefont {Braden}\ \bibnamefont
  {Kronheim}}, \bibinfo {author} {\bibfnamefont {Michelle}\ \bibnamefont
  {Kuchera}}, \bibinfo {author} {\bibfnamefont {Harrison}\ \bibnamefont
  {Prosper}}, \ and\ \bibinfo {author} {\bibfnamefont {Alexander}\ \bibnamefont
  {Karbo}},\ }\bibfield  {title} {\enquote {\bibinfo {title} {{Bayesian Neural
  Networks for Fast SUSY Predictions}},}\ }\href {\doibase
  10.1016/j.physletb.2020.136041} {\  (\bibinfo {year} {2020}),\
  10.1016/j.physletb.2020.136041},\ \Eprint {http://arxiv.org/abs/2007.04506}
  {arXiv:2007.04506 [hep-ph]} \BibitemShut {NoStop}%
\bibitem [{\citenamefont {Araz}\ and\ \citenamefont
  {Spannowsky}(2021)}]{Araz:2021wqm}%
  \BibitemOpen
  \bibfield  {author} {\bibinfo {author} {\bibfnamefont {Jack~Y.}\ \bibnamefont
  {Araz}}\ and\ \bibinfo {author} {\bibfnamefont {Michael}\ \bibnamefont
  {Spannowsky}},\ }\bibfield  {title} {\enquote {\bibinfo {title} {{Combine and
  Conquer: Event Reconstruction with Bayesian Ensemble Neural Networks}},}\
  }\href@noop {} {\  (\bibinfo {year} {2021})},\ \Eprint
  {http://arxiv.org/abs/2102.01078} {arXiv:2102.01078 [hep-ph]} \BibitemShut
  {NoStop}%
\bibitem [{\citenamefont {Kronheim}\ \emph {et~al.}(2021)\citenamefont
  {Kronheim}, \citenamefont {Kuchera}, \citenamefont {Prosper},\ and\
  \citenamefont {Ramanujan}}]{Kronheim:2021hdb}%
  \BibitemOpen
  \bibfield  {author} {\bibinfo {author} {\bibfnamefont {Braden}\ \bibnamefont
  {Kronheim}}, \bibinfo {author} {\bibfnamefont {Michelle~P.}\ \bibnamefont
  {Kuchera}}, \bibinfo {author} {\bibfnamefont {Harrison~B.}\ \bibnamefont
  {Prosper}}, \ and\ \bibinfo {author} {\bibfnamefont {Raghuram}\ \bibnamefont
  {Ramanujan}},\ }\bibfield  {title} {\enquote {\bibinfo {title} {{Implicit
  Quantile Neural Networks for Jet Simulation and Correction}},}\ }\href@noop
  {} {\  (\bibinfo {year} {2021})},\ \Eprint {http://arxiv.org/abs/2111.11415}
  {arXiv:2111.11415 [physics.comp-ph]} \BibitemShut {NoStop}%
\bibitem [{\citenamefont {Dalmasso}\ \emph {et~al.}(2021)\citenamefont
  {Dalmasso}, \citenamefont {Zhao}, \citenamefont {Izbicki},\ and\
  \citenamefont {Lee}}]{https://doi.org/10.48550/arxiv.2107.03920}%
  \BibitemOpen
  \bibfield  {author} {\bibinfo {author} {\bibfnamefont {Niccolò}\
  \bibnamefont {Dalmasso}}, \bibinfo {author} {\bibfnamefont {David}\
  \bibnamefont {Zhao}}, \bibinfo {author} {\bibfnamefont {Rafael}\ \bibnamefont
  {Izbicki}}, \ and\ \bibinfo {author} {\bibfnamefont {Ann~B.}\ \bibnamefont
  {Lee}},\ }\href {\doibase 10.48550/ARXIV.2107.03920} {\enquote {\bibinfo
  {title} {Likelihood-free frequentist inference: Bridging classical statistics
  and machine learning in simulation and uncertainty quantification},}\ }
  (\bibinfo {year} {2021})\BibitemShut {NoStop}%
\bibitem [{\citenamefont {Belghazi}\ \emph {et~al.}(2018)\citenamefont
  {Belghazi}, \citenamefont {Baratin}, \citenamefont {Rajeswar}, \citenamefont
  {Ozair}, \citenamefont {Bengio}, \citenamefont {Courville},\ and\
  \citenamefont {Hjelm}}]{belghazi2018mine}%
  \BibitemOpen
  \bibfield  {author} {\bibinfo {author} {\bibfnamefont {Mohamed~Ishmael}\
  \bibnamefont {Belghazi}}, \bibinfo {author} {\bibfnamefont {Aristide}\
  \bibnamefont {Baratin}}, \bibinfo {author} {\bibfnamefont {Sai}\ \bibnamefont
  {Rajeswar}}, \bibinfo {author} {\bibfnamefont {Sherjil}\ \bibnamefont
  {Ozair}}, \bibinfo {author} {\bibfnamefont {Yoshua}\ \bibnamefont {Bengio}},
  \bibinfo {author} {\bibfnamefont {Aaron}\ \bibnamefont {Courville}}, \ and\
  \bibinfo {author} {\bibfnamefont {R~Devon}\ \bibnamefont {Hjelm}},\
  }\href@noop {} {\enquote {\bibinfo {title} {Mine: Mutual information neural
  estimation},}\ } (\bibinfo {year} {2018}),\ \Eprint
  {http://arxiv.org/abs/1801.04062} {arXiv:1801.04062 [cs.LG]} \BibitemShut
  {NoStop}%
\bibitem [{\citenamefont {Donsker}\ and\ \citenamefont
  {Varadhan}(1975)}]{Donsker1975AsymptoticEO}%
  \BibitemOpen
  \bibfield  {author} {\bibinfo {author} {\bibfnamefont {Monroe~D.}\
  \bibnamefont {Donsker}}\ and\ \bibinfo {author} {\bibfnamefont {S.~R.~S.}\
  \bibnamefont {Varadhan}},\ }\bibfield  {title} {\enquote {\bibinfo {title}
  {Asymptotic evaluation of certain markov process expectations for large
  time},}\ \ }(\bibinfo {year} {1975})\BibitemShut {NoStop}%
\bibitem [{\citenamefont {Kullback}\ and\ \citenamefont
  {Leibler}(1951)}]{kullback1951information}%
  \BibitemOpen
  \bibfield  {author} {\bibinfo {author} {\bibfnamefont {Solomon}\ \bibnamefont
  {Kullback}}\ and\ \bibinfo {author} {\bibfnamefont {Richard~A}\ \bibnamefont
  {Leibler}},\ }\bibfield  {title} {\enquote {\bibinfo {title} {On information
  and sufficiency},}\ }\href@noop {} {\bibfield  {journal} {\bibinfo  {journal}
  {The annals of mathematical statistics}\ }\textbf {\bibinfo {volume} {22}},\
  \bibinfo {pages} {79--86} (\bibinfo {year} {1951})}\BibitemShut {NoStop}%
\bibitem [{\citenamefont {Donsker}\ and\ \citenamefont
  {Varadhan}(1976)}]{e5879cd3d84b462abf51f06791e5ba28}%
  \BibitemOpen
  \bibfield  {author} {\bibinfo {author} {\bibfnamefont {{M. D.}}\ \bibnamefont
  {Donsker}}\ and\ \bibinfo {author} {\bibfnamefont {{S. R.S.}}\ \bibnamefont
  {Varadhan}},\ }\bibfield  {title} {\enquote {\bibinfo {title} {Asymptotic
  evaluation of certain markov process expectations for large time—iii},}\
  }\href {\doibase 10.1002/cpa.3160290405} {\bibfield  {journal} {\bibinfo
  {journal} {Communications on Pure and Applied Mathematics}\ }\textbf
  {\bibinfo {volume} {29}},\ \bibinfo {pages} {389--461} (\bibinfo {year}
  {1976})},\ \bibinfo {note} {copyright: Copyright 2016 Elsevier B.V., All
  rights reserved.}\BibitemShut {Stop}%
\bibitem [{\citenamefont {Nowozin}\ \emph {et~al.}(2016)\citenamefont
  {Nowozin}, \citenamefont {Cseke},\ and\ \citenamefont
  {Tomioka}}]{nowozin2016fgan}%
  \BibitemOpen
  \bibfield  {author} {\bibinfo {author} {\bibfnamefont {Sebastian}\
  \bibnamefont {Nowozin}}, \bibinfo {author} {\bibfnamefont {Botond}\
  \bibnamefont {Cseke}}, \ and\ \bibinfo {author} {\bibfnamefont {Ryota}\
  \bibnamefont {Tomioka}},\ }\href@noop {} {\enquote {\bibinfo {title} {f-gan:
  Training generative neural samplers using variational divergence
  minimization},}\ } (\bibinfo {year} {2016}),\ \Eprint
  {http://arxiv.org/abs/1606.00709} {arXiv:1606.00709 [stat.ML]} \BibitemShut
  {NoStop}%
\bibitem [{\citenamefont {Nguyen}\ \emph {et~al.}(2010)\citenamefont {Nguyen},
  \citenamefont {Wainwright},\ and\ \citenamefont {Jordan}}]{Nguyen_2010}%
  \BibitemOpen
  \bibfield  {author} {\bibinfo {author} {\bibfnamefont {XuanLong}\
  \bibnamefont {Nguyen}}, \bibinfo {author} {\bibfnamefont {Martin~J.}\
  \bibnamefont {Wainwright}}, \ and\ \bibinfo {author} {\bibfnamefont
  {Michael~I.}\ \bibnamefont {Jordan}},\ }\bibfield  {title} {\enquote
  {\bibinfo {title} {Estimating divergence functionals and the likelihood ratio
  by convex risk minimization},}\ }\href {\doibase 10.1109/tit.2010.2068870}
  {\bibfield  {journal} {\bibinfo  {journal} {IEEE Transactions on Information
  Theory}\ }\textbf {\bibinfo {volume} {56}},\ \bibinfo {pages} {5847–5861}
  (\bibinfo {year} {2010})}\BibitemShut {NoStop}%
\bibitem [{\citenamefont {D'Agnolo}\ and\ \citenamefont
  {Wulzer}(2019)}]{DAgnolo:2018cun}%
  \BibitemOpen
  \bibfield  {author} {\bibinfo {author} {\bibfnamefont {Raffaele~Tito}\
  \bibnamefont {D'Agnolo}}\ and\ \bibinfo {author} {\bibfnamefont {Andrea}\
  \bibnamefont {Wulzer}},\ }\bibfield  {title} {\enquote {\bibinfo {title}
  {{Learning New Physics from a Machine}},}\ }\href {\doibase
  10.1103/PhysRevD.99.015014} {\bibfield  {journal} {\bibinfo  {journal} {Phys.
  Rev. D}\ }\textbf {\bibinfo {volume} {99}},\ \bibinfo {pages} {015014}
  (\bibinfo {year} {2019})},\ \Eprint {http://arxiv.org/abs/1806.02350}
  {arXiv:1806.02350 [hep-ph]} \BibitemShut {NoStop}%
\bibitem [{\citenamefont {D'Agnolo}\ \emph {et~al.}(2019)\citenamefont
  {D'Agnolo}, \citenamefont {Grosso}, \citenamefont {Pierini}, \citenamefont
  {Wulzer},\ and\ \citenamefont {Zanetti}}]{DAgnolo:2019vbw}%
  \BibitemOpen
  \bibfield  {author} {\bibinfo {author} {\bibfnamefont {Raffaele~Tito}\
  \bibnamefont {D'Agnolo}}, \bibinfo {author} {\bibfnamefont {Gaia}\
  \bibnamefont {Grosso}}, \bibinfo {author} {\bibfnamefont {Maurizio}\
  \bibnamefont {Pierini}}, \bibinfo {author} {\bibfnamefont {Andrea}\
  \bibnamefont {Wulzer}}, \ and\ \bibinfo {author} {\bibfnamefont {Marco}\
  \bibnamefont {Zanetti}},\ }\bibfield  {title} {\enquote {\bibinfo {title}
  {{Learning Multivariate New Physics}},}\ }\href@noop {} {\  (\bibinfo {year}
  {2019})},\ \Eprint {http://arxiv.org/abs/1912.12155} {arXiv:1912.12155
  [hep-ph]} \BibitemShut {NoStop}%
\bibitem [{\citenamefont {Nachman}\ and\ \citenamefont
  {Thaler}(2021)}]{nachman2021e}%
  \BibitemOpen
  \bibfield  {author} {\bibinfo {author} {\bibfnamefont {Benjamin}\
  \bibnamefont {Nachman}}\ and\ \bibinfo {author} {\bibfnamefont {Jesse}\
  \bibnamefont {Thaler}},\ }\href@noop {} {\enquote {\bibinfo {title} {E
  pluribus unum ex machina: Learning from many collider events at once},}\ }
  (\bibinfo {year} {2021}),\ \Eprint {http://arxiv.org/abs/2101.07263}
  {arXiv:2101.07263 [physics.data-an]} \BibitemShut {NoStop}%
\bibitem [{\citenamefont {Ruderman}\ \emph {et~al.}(2012)\citenamefont
  {Ruderman}, \citenamefont {Reid}, \citenamefont {Garcia-Garcia},\ and\
  \citenamefont {Petterson}}]{ruderman2012tighter}%
  \BibitemOpen
  \bibfield  {author} {\bibinfo {author} {\bibfnamefont {Avraham}\ \bibnamefont
  {Ruderman}}, \bibinfo {author} {\bibfnamefont {Mark}\ \bibnamefont {Reid}},
  \bibinfo {author} {\bibfnamefont {Dario}\ \bibnamefont {Garcia-Garcia}}, \
  and\ \bibinfo {author} {\bibfnamefont {James}\ \bibnamefont {Petterson}},\
  }\href@noop {} {\enquote {\bibinfo {title} {Tighter variational
  representations of f-divergences via restriction to probability measures},}\
  } (\bibinfo {year} {2012}),\ \Eprint {http://arxiv.org/abs/1206.4664}
  {arXiv:1206.4664 [cs.LG]} \BibitemShut {NoStop}%
\bibitem [{\citenamefont {Khachatryan}\ \emph
  {et~al.}(2017{\natexlab{a}})\citenamefont {Khachatryan}, \citenamefont
  {Sirunyan}, \citenamefont {Tumasyan}, \citenamefont {Adam}, \citenamefont
  {Asilar}, \citenamefont {Bergauer}, \citenamefont {Brandstetter},
  \citenamefont {Brondolin}, \citenamefont {Dragicevic}, \citenamefont {Erö},\
  and\ \citenamefont {et~al.}}]{Khachatryan_2017}%
  \BibitemOpen
  \bibfield  {author} {\bibinfo {author} {\bibfnamefont {V.}~\bibnamefont
  {Khachatryan}}, \bibinfo {author} {\bibfnamefont {A.M.}\ \bibnamefont
  {Sirunyan}}, \bibinfo {author} {\bibfnamefont {A.}~\bibnamefont {Tumasyan}},
  \bibinfo {author} {\bibfnamefont {W.}~\bibnamefont {Adam}}, \bibinfo {author}
  {\bibfnamefont {E.}~\bibnamefont {Asilar}}, \bibinfo {author} {\bibfnamefont
  {T.}~\bibnamefont {Bergauer}}, \bibinfo {author} {\bibfnamefont
  {J.}~\bibnamefont {Brandstetter}}, \bibinfo {author} {\bibfnamefont
  {E.}~\bibnamefont {Brondolin}}, \bibinfo {author} {\bibfnamefont
  {M.}~\bibnamefont {Dragicevic}}, \bibinfo {author} {\bibfnamefont
  {J.}~\bibnamefont {Erö}}, \ and\ \bibinfo {author} {\bibnamefont {et~al.}},\
  }\bibfield  {title} {\enquote {\bibinfo {title} {Jet energy scale and
  resolution in the cms experiment in pp collisions at 8 tev},}\ }\href
  {\doibase 10.1088/1748-0221/12/02/p02014} {\bibfield  {journal} {\bibinfo
  {journal} {Journal of Instrumentation}\ }\textbf {\bibinfo {volume} {12}},\
  \bibinfo {pages} {P02014–P02014} (\bibinfo {year}
  {2017}{\natexlab{a}})}\BibitemShut {NoStop}%
\bibitem [{\citenamefont {collaboration}(2011)}]{2011jes}%
  \BibitemOpen
  \bibfield  {author} {\bibinfo {author} {\bibfnamefont {The~CMS}\ \bibnamefont
  {collaboration}},\ }\bibfield  {title} {\enquote {\bibinfo {title}
  {Determination of jet energy calibration and transverse momentum resolution
  in cms},}\ }\href {\doibase 10.1088/1748-0221/6/11/p11002} {\bibfield
  {journal} {\bibinfo  {journal} {Journal of Instrumentation}\ }\textbf
  {\bibinfo {volume} {6}},\ \bibinfo {pages} {P11002–P11002} (\bibinfo {year}
  {2011})}\BibitemShut {NoStop}%
\bibitem [{\citenamefont {Khachatryan}\ \emph
  {et~al.}(2017{\natexlab{b}})\citenamefont {Khachatryan} \emph
  {et~al.}}]{CMS:2016lmd}%
  \BibitemOpen
  \bibfield  {author} {\bibinfo {author} {\bibfnamefont {Vardan}\ \bibnamefont
  {Khachatryan}} \emph {et~al.} (\bibinfo {collaboration} {CMS}),\ }\bibfield
  {title} {\enquote {\bibinfo {title} {{Jet energy scale and resolution in the
  CMS experiment in pp collisions at 8 TeV}},}\ }\href {\doibase
  10.1088/1748-0221/12/02/P02014} {\bibfield  {journal} {\bibinfo  {journal}
  {JINST}\ }\textbf {\bibinfo {volume} {12}},\ \bibinfo {pages} {P02014}
  (\bibinfo {year} {2017}{\natexlab{b}})},\ \Eprint
  {http://arxiv.org/abs/1607.03663} {arXiv:1607.03663 [hep-ex]} \BibitemShut
  {NoStop}%
\bibitem [{\citenamefont {Aaboud}\ \emph {et~al.}(2017)\citenamefont {Aaboud}
  \emph {et~al.}}]{ATLAS:2017bje}%
  \BibitemOpen
  \bibfield  {author} {\bibinfo {author} {\bibfnamefont {M.}~\bibnamefont
  {Aaboud}} \emph {et~al.} (\bibinfo {collaboration} {ATLAS}),\ }\bibfield
  {title} {\enquote {\bibinfo {title} {{Jet energy scale measurements and their
  systematic uncertainties in proton-proton collisions at $\sqrt{s} = 13$ TeV
  with the ATLAS detector}},}\ }\href {\doibase 10.1103/PhysRevD.96.072002}
  {\bibfield  {journal} {\bibinfo  {journal} {Phys. Rev. D}\ }\textbf {\bibinfo
  {volume} {96}},\ \bibinfo {pages} {072002} (\bibinfo {year} {2017})},\
  \Eprint {http://arxiv.org/abs/1703.09665} {arXiv:1703.09665 [hep-ex]}
  \BibitemShut {NoStop}%
\bibitem [{\citenamefont {Aad}\ \emph {et~al.}(2015)\citenamefont {Aad} \emph
  {et~al.}}]{ATLAS:2014hvo}%
  \BibitemOpen
  \bibfield  {author} {\bibinfo {author} {\bibfnamefont {Georges}\ \bibnamefont
  {Aad}} \emph {et~al.} (\bibinfo {collaboration} {ATLAS}),\ }\bibfield
  {title} {\enquote {\bibinfo {title} {{Jet energy measurement and its
  systematic uncertainty in proton-proton collisions at $\sqrt{s}=7$ TeV with
  the ATLAS detector}},}\ }\href {\doibase 10.1140/epjc/s10052-014-3190-y}
  {\bibfield  {journal} {\bibinfo  {journal} {Eur. Phys. J. C}\ }\textbf
  {\bibinfo {volume} {75}},\ \bibinfo {pages} {17} (\bibinfo {year} {2015})},\
  \Eprint {http://arxiv.org/abs/1406.0076} {arXiv:1406.0076 [hep-ex]}
  \BibitemShut {NoStop}%
\bibitem [{\citenamefont {Aaboud}\ \emph {et~al.}(2020)\citenamefont {Aaboud}
  \emph {et~al.}}]{ATLAS:2019oxp}%
  \BibitemOpen
  \bibfield  {author} {\bibinfo {author} {\bibfnamefont {Morad}\ \bibnamefont
  {Aaboud}} \emph {et~al.} (\bibinfo {collaboration} {ATLAS}),\ }\bibfield
  {title} {\enquote {\bibinfo {title} {{Determination of jet calibration and
  energy resolution in proton-proton collisions at $\sqrt{s}$ = 8 TeV using the
  ATLAS detector}},}\ }\href {\doibase 10.1140/epjc/s10052-020-08477-8}
  {\bibfield  {journal} {\bibinfo  {journal} {Eur. Phys. J. C}\ }\textbf
  {\bibinfo {volume} {80}},\ \bibinfo {pages} {1104} (\bibinfo {year}
  {2020})},\ \Eprint {http://arxiv.org/abs/1910.04482} {arXiv:1910.04482
  [hep-ex]} \BibitemShut {NoStop}%
\bibitem [{\citenamefont {Cukierman}\ and\ \citenamefont
  {Nachman}(2017)}]{Cukierman_2017}%
  \BibitemOpen
  \bibfield  {author} {\bibinfo {author} {\bibfnamefont {Aviv}\ \bibnamefont
  {Cukierman}}\ and\ \bibinfo {author} {\bibfnamefont {Benjamin}\ \bibnamefont
  {Nachman}},\ }\bibfield  {title} {\enquote {\bibinfo {title} {Mathematical
  properties of numerical inversion for jet calibrations},}\ }\href {\doibase
  10.1016/j.nima.2017.03.038} {\bibfield  {journal} {\bibinfo  {journal}
  {Nuclear Instruments and Methods in Physics Research Section A: Accelerators,
  Spectrometers, Detectors and Associated Equipment}\ }\textbf {\bibinfo
  {volume} {858}},\ \bibinfo {pages} {1–11} (\bibinfo {year}
  {2017})}\BibitemShut {NoStop}%
\bibitem [{\citenamefont {Rao}(2016)}]{cmspressrelease}%
  \BibitemOpen
  \bibfield  {author} {\bibinfo {author} {\bibfnamefont {Achintya}\
  \bibnamefont {Rao}},\ }\bibfield  {title} {\enquote {\bibinfo {title} {Cms
  releases new batch of lhc open data},}\ }\href
  {https://home.cern/news/news/computing/cms-releases-new-batch-lhc-open-data}
  {\  (\bibinfo {year} {2016})}\BibitemShut {NoStop}%
\bibitem [{cer()}]{cernopendata}%
  \BibitemOpen
  \href {http://opendata.cern.ch} {\enquote {\bibinfo {title} {Cern open data
  portal},}\ }\BibitemShut {NoStop}%
\bibitem [{\citenamefont {Komiske}\ \emph {et~al.}(2020)\citenamefont
  {Komiske}, \citenamefont {Mastandrea}, \citenamefont {Metodiev},
  \citenamefont {Naik},\ and\ \citenamefont {Thaler}}]{Komiske_2020}%
  \BibitemOpen
  \bibfield  {author} {\bibinfo {author} {\bibfnamefont {Patrick~T.}\
  \bibnamefont {Komiske}}, \bibinfo {author} {\bibfnamefont {Radha}\
  \bibnamefont {Mastandrea}}, \bibinfo {author} {\bibfnamefont {Eric~M.}\
  \bibnamefont {Metodiev}}, \bibinfo {author} {\bibfnamefont {Preksha}\
  \bibnamefont {Naik}}, \ and\ \bibinfo {author} {\bibfnamefont {Jesse}\
  \bibnamefont {Thaler}},\ }\bibfield  {title} {\enquote {\bibinfo {title}
  {Exploring the space of jets with cms open data},}\ }\href {\doibase
  10.1103/physrevd.101.034009} {\bibfield  {journal} {\bibinfo  {journal}
  {Physical Review D}\ }\textbf {\bibinfo {volume} {101}} (\bibinfo {year}
  {2020}),\ 10.1103/physrevd.101.034009}\BibitemShut {NoStop}%
\bibitem [{\citenamefont {Sjöstrand}\ \emph {et~al.}(2006)\citenamefont
  {Sjöstrand}, \citenamefont {Mrenna},\ and\ \citenamefont
  {Skands}}]{Sj_strand_2006}%
  \BibitemOpen
  \bibfield  {author} {\bibinfo {author} {\bibfnamefont {Torbjörn}\
  \bibnamefont {Sjöstrand}}, \bibinfo {author} {\bibfnamefont {Stephen}\
  \bibnamefont {Mrenna}}, \ and\ \bibinfo {author} {\bibfnamefont {Peter}\
  \bibnamefont {Skands}},\ }\bibfield  {title} {\enquote {\bibinfo {title}
  {{PYTHIA} 6.4 physics and manual},}\ }\href {\doibase
  10.1088/1126-6708/2006/05/026} {\bibfield  {journal} {\bibinfo  {journal}
  {Journal of High Energy Physics}\ }\textbf {\bibinfo {volume} {2006}},\
  \bibinfo {pages} {026--026} (\bibinfo {year} {2006})}\BibitemShut {NoStop}%
\bibitem [{\citenamefont {Agostinelli}\ \emph {et~al.}(2003)\citenamefont
  {Agostinelli} \emph {et~al.}}]{AGOSTINELLI2003250}%
  \BibitemOpen
  \bibfield  {author} {\bibinfo {author} {\bibfnamefont {S.}~\bibnamefont
  {Agostinelli}} \emph {et~al.},\ }\bibfield  {title} {\enquote {\bibinfo
  {title} {Geant4—a simulation toolkit},}\ }\href {\doibase
  https://doi.org/10.1016/S0168-9002(03)01368-8} {\bibfield  {journal}
  {\bibinfo  {journal} {Nuclear Instruments and Methods in Physics Research
  Section A: Accelerators, Spectrometers, Detectors and Associated Equipment}\
  }\textbf {\bibinfo {volume} {506}},\ \bibinfo {pages} {250--303} (\bibinfo
  {year} {2003})}\BibitemShut {NoStop}%
\bibitem [{\citenamefont {Komiske}\ \emph
  {et~al.}(2019{\natexlab{a}})\citenamefont {Komiske}, \citenamefont
  {Mastandrea}, \citenamefont {Metodiev}, \citenamefont {Naik},\ and\
  \citenamefont {Thaler}}]{komiske_patrick_2019_3340205}%
  \BibitemOpen
  \bibfield  {author} {\bibinfo {author} {\bibfnamefont {Patrick}\ \bibnamefont
  {Komiske}}, \bibinfo {author} {\bibfnamefont {Radha}\ \bibnamefont
  {Mastandrea}}, \bibinfo {author} {\bibfnamefont {Eric}\ \bibnamefont
  {Metodiev}}, \bibinfo {author} {\bibfnamefont {Preksha}\ \bibnamefont
  {Naik}}, \ and\ \bibinfo {author} {\bibfnamefont {Jesse}\ \bibnamefont
  {Thaler}},\ }\href {\doibase 10.5281/zenodo.3340205} {\enquote {\bibinfo
  {title} {{CMS 2011A Open Data | Jet Primary Dataset | pT > 375 GeV | MOD HDF5
  Format}},}\ } (\bibinfo {year} {2019}{\natexlab{a}})\BibitemShut {NoStop}%
\bibitem [{\citenamefont {Cacciari}\ and\ \citenamefont
  {Salam}(2006)}]{Cacciari:2005hq}%
  \BibitemOpen
  \bibfield  {author} {\bibinfo {author} {\bibfnamefont {Matteo}\ \bibnamefont
  {Cacciari}}\ and\ \bibinfo {author} {\bibfnamefont {Gavin~P.}\ \bibnamefont
  {Salam}},\ }\bibfield  {title} {\enquote {\bibinfo {title} {{Dispelling the
  $N^{3}$ myth for the $k_t$ jet-finder}},}\ }\href {\doibase
  10.1016/j.physletb.2006.08.037} {\bibfield  {journal} {\bibinfo  {journal}
  {Phys. Lett.}\ }\textbf {\bibinfo {volume} {B641}},\ \bibinfo {pages} {57}
  (\bibinfo {year} {2006})},\ \Eprint {http://arxiv.org/abs/hep-ph/0512210}
  {arXiv:hep-ph/0512210 [hep-ph]} \BibitemShut {NoStop}%
\bibitem [{\citenamefont {Cacciari}\ \emph {et~al.}(2008)\citenamefont
  {Cacciari}, \citenamefont {Salam},\ and\ \citenamefont
  {Soyez}}]{Cacciari_2008}%
  \BibitemOpen
  \bibfield  {author} {\bibinfo {author} {\bibfnamefont {Matteo}\ \bibnamefont
  {Cacciari}}, \bibinfo {author} {\bibfnamefont {Gavin~P}\ \bibnamefont
  {Salam}}, \ and\ \bibinfo {author} {\bibfnamefont {Gregory}\ \bibnamefont
  {Soyez}},\ }\bibfield  {title} {\enquote {\bibinfo {title} {The anti-ktjet
  clustering algorithm},}\ }\href {\doibase 10.1088/1126-6708/2008/04/063}
  {\bibfield  {journal} {\bibinfo  {journal} {Journal of High Energy Physics}\
  }\textbf {\bibinfo {volume} {2008}},\ \bibinfo {pages} {063–063} (\bibinfo
  {year} {2008})}\BibitemShut {NoStop}%
\bibitem [{\citenamefont {Cacciari}\ \emph {et~al.}(2012)\citenamefont
  {Cacciari}, \citenamefont {Salam},\ and\ \citenamefont
  {Soyez}}]{Cacciari:2011ma}%
  \BibitemOpen
  \bibfield  {author} {\bibinfo {author} {\bibfnamefont {Matteo}\ \bibnamefont
  {Cacciari}}, \bibinfo {author} {\bibfnamefont {Gavin~P.}\ \bibnamefont
  {Salam}}, \ and\ \bibinfo {author} {\bibfnamefont {Gregory}\ \bibnamefont
  {Soyez}},\ }\bibfield  {title} {\enquote {\bibinfo {title} {{FastJet User
  Manual}},}\ }\href {\doibase 10.1140/epjc/s10052-012-1896-2} {\bibfield
  {journal} {\bibinfo  {journal} {Eur. Phys. J.}\ }\textbf {\bibinfo {volume}
  {C72}},\ \bibinfo {pages} {1896} (\bibinfo {year} {2012})},\ \Eprint
  {http://arxiv.org/abs/1111.6097} {arXiv:1111.6097 [hep-ph]} \BibitemShut
  {NoStop}%
\bibitem [{CMS(2010)}]{CMS:2010xta}%
  \BibitemOpen
  \bibfield  {title} {\enquote {\bibinfo {title} {{Jet Performance in pp
  Collisions at 7 TeV}},}\ }\href@noop {} {\  (\bibinfo {year}
  {2010})}\BibitemShut {NoStop}%
\bibitem [{\citenamefont {Komiske}\ \emph
  {et~al.}(2019{\natexlab{b}})\citenamefont {Komiske}, \citenamefont
  {Metodiev},\ and\ \citenamefont {Thaler}}]{Komiske_2019}%
  \BibitemOpen
  \bibfield  {author} {\bibinfo {author} {\bibfnamefont {Patrick~T.}\
  \bibnamefont {Komiske}}, \bibinfo {author} {\bibfnamefont {Eric~M.}\
  \bibnamefont {Metodiev}}, \ and\ \bibinfo {author} {\bibfnamefont {Jesse}\
  \bibnamefont {Thaler}},\ }\bibfield  {title} {\enquote {\bibinfo {title}
  {Energy flow networks: deep sets for particle jets},}\ }\href {\doibase
  10.1007/jhep01(2019)121} {\bibfield  {journal} {\bibinfo  {journal} {Journal
  of High Energy Physics}\ }\textbf {\bibinfo {volume} {2019}} (\bibinfo {year}
  {2019}{\natexlab{b}}),\ 10.1007/jhep01(2019)121}\BibitemShut {NoStop}%
\bibitem [{\citenamefont {Zaheer}\ \emph {et~al.}(2017)\citenamefont {Zaheer},
  \citenamefont {Kottur}, \citenamefont {Ravanbakhsh}, \citenamefont {Poczos},
  \citenamefont {Salakhutdinov},\ and\ \citenamefont
  {Smola}}]{NIPS2017_f22e4747}%
  \BibitemOpen
  \bibfield  {author} {\bibinfo {author} {\bibfnamefont {Manzil}\ \bibnamefont
  {Zaheer}}, \bibinfo {author} {\bibfnamefont {Satwik}\ \bibnamefont {Kottur}},
  \bibinfo {author} {\bibfnamefont {Siamak}\ \bibnamefont {Ravanbakhsh}},
  \bibinfo {author} {\bibfnamefont {Barnabas}\ \bibnamefont {Poczos}}, \bibinfo
  {author} {\bibfnamefont {Russ~R}\ \bibnamefont {Salakhutdinov}}, \ and\
  \bibinfo {author} {\bibfnamefont {Alexander~J}\ \bibnamefont {Smola}},\
  }\bibfield  {title} {\enquote {\bibinfo {title} {Deep sets},}\ }in\ \href
  {https://proceedings.neurips.cc/paper/2017/file/f22e4747da1aa27e363d86d40ff442fe-Paper.pdf}
  {\emph {\bibinfo {booktitle} {Advances in Neural Information Processing
  Systems}}},\ Vol.~\bibinfo {volume} {30},\ \bibinfo {editor} {edited by\
  \bibinfo {editor} {\bibfnamefont {I.}~\bibnamefont {Guyon}}, \bibinfo
  {editor} {\bibfnamefont {U.~V.}\ \bibnamefont {Luxburg}}, \bibinfo {editor}
  {\bibfnamefont {S.}~\bibnamefont {Bengio}}, \bibinfo {editor} {\bibfnamefont
  {H.}~\bibnamefont {Wallach}}, \bibinfo {editor} {\bibfnamefont
  {R.}~\bibnamefont {Fergus}}, \bibinfo {editor} {\bibfnamefont
  {S.}~\bibnamefont {Vishwanathan}}, \ and\ \bibinfo {editor} {\bibfnamefont
  {R.}~\bibnamefont {Garnett}}}\ (\bibinfo  {publisher} {Curran Associates,
  Inc.},\ \bibinfo {year} {2017})\BibitemShut {NoStop}%
\bibitem [{\citenamefont {Kingma}\ and\ \citenamefont
  {Ba}(2017)}]{kingma2017adam}%
  \BibitemOpen
  \bibfield  {author} {\bibinfo {author} {\bibfnamefont {Diederik~P.}\
  \bibnamefont {Kingma}}\ and\ \bibinfo {author} {\bibfnamefont {Jimmy}\
  \bibnamefont {Ba}},\ }\href@noop {} {\enquote {\bibinfo {title} {Adam: A
  method for stochastic optimization},}\ } (\bibinfo {year} {2017}),\ \Eprint
  {http://arxiv.org/abs/1412.6980} {arXiv:1412.6980 [cs.LG]} \BibitemShut
  {NoStop}%
\end{thebibliography}%

\end{document}